\documentclass[twocolumn,times]{aastex701}
\usepackage{graphicx} 
\usepackage{subcaption}
\usepackage{url}
\usepackage{dirtytalk}
\usepackage{xcolor}

\usepackage{float}
\usepackage{amsmath}

\newcommand{\xiiico}[0]{$^{13}$CO}
\newcommand{\xiico}[0]{$^{12}$CO}
\newcommand{\xiiicotwoone}[0]{$^{13}$CO(2--1)}
\newcommand{\xiicotwoone}[0]{$^{12}$CO(2--1)}
\newcommand{\kms}{\,km\,s$^{-1}$}
\shorttitle{TIME Galactic Center}
\shortauthors{Yang et al.}
\begin{document}
\title{TIME Commissioning Observations: I. Mapping Dust and Molecular Gas in the Sgr A Molecular Cloud Complex at the Galactic Center}
\author[0009-0002-7777-2351]{Selina F. Yang}
\affiliation{Department of Physics, Cornell University, Ithaca NY 14853, USA}
\email{fy235@cornell.edu}
\author[0009-0002-1152-4953]{Sophie M.~McAtee}
\affiliation{Department of Physics, Cornell University, Ithaca NY 14853, USA}
\email{smm549@cornell.edu}
\author[0000-0002-9813-0270]{Benjamin J.~Vaughan}
\affiliation{Department of Physics, Cornell University, Ithaca NY 14853, USA}
\email{bjv37@cornell.edu}
\author[0009-0005-4099-8842]{Abigail T.~Crites}
\affiliation{Department of Physics, Cornell University, Ithaca NY 14853, USA}
\affiliation{Department of Astronomy, Cornell University, Ithaca NY 14853, USA}
\email{atc72@cornell.edu}
\affiliation{Department of Physics, California Institute of Technology, Pasadena, California 91125, USA}
\author[0000-0002-0941-0407]{Victoria L.~Butler}
\affiliation{Department of Physics, Cornell University, Ithaca NY 14853, USA}
\email{vlb59@cornell.edu}
\author[0000-0003-2618-6504]{Dongwoo T.~Chung}
\affiliation{Department of Astronomy, Cornell University, Ithaca NY 14853, USA}
\email{dongwooc@cornell.edu}
\author{Ryan P.~Keenan}
\affiliation{Max-Planck-Institut f\"{u}r Astronomie, K\"{o}nigstuhl 17, D-69117 Heidelberg, Germany}
\email{keenan@mpia.de}
\author{Dang Pham}
\affiliation{JILA and Department of Astrophysical and Planetary Sciences, CU Boulder, Boulder, CO 80309, USA}
\email{dang.pham@astro.utoronto.ca}
\affiliation{David A.~Dunlap Department of Astronomy and Astrophysics, University of Toronto, 50 St. George Street, Toronto, ON M5S 3H4, Canada}
\author[0009-0004-1047-3714]{Shwetha Prakash}
\affiliation{Department of Physics, Cornell University, Ithaca NY 14853, USA}
\email{sp2358@cornell.edu}
\author{James~J.~Bock}
\affiliation{Department of Physics, California Institute of Technology, Pasadena, California 91125, USA}
\affiliation{Jet Propulsion Laboratory, California Institute of Technology, Pasadena, California 91109, USA}
\email{jjb@astro.caltech.edu}
\author{Charles~M.~Bradford}
\affiliation{Department of Physics, California Institute of Technology, Pasadena, California 91125, USA}
\affiliation{Jet Propulsion Laboratory, California Institute of Technology, Pasadena, California 91109, USA}
\email{bradford@submm.caltech.edu}
\author{Tzu-Ching~Chang}
\affiliation{Jet Propulsion Laboratory, California Institute of Technology, Pasadena, California 91109, USA}
\email{tzu-ching.chang@jpl.nasa.gov}
\author{Yun-Ting~Cheng}
\affiliation{Department of Physics, California Institute of Technology, Pasadena, California 91125, USA}
\affiliation{Jet Propulsion Laboratory, California Institute of Technology, Pasadena, California 91109, USA}
\email{ycheng3@caltech.edu}
\author[0009-0007-6638-5774]{Audrey~Dunn}
\affiliation{Rochester Institute of Technology, Rochester, NY, 14623, USA}
\email{akd5648@g.rit.edu}
\author{Nicholas~Emerson}
\affiliation{Department of Astronomy and Steward Observatory, University of Arizona, 933 N Cherry Avenue, Tucson, AZ 85721, USA}
\email{nemerson@arizona.edu}
\author{Clifford~Frez}
\affiliation{Jet Propulsion Laboratory, California Institute of Technology, Pasadena, California 91109, USA}
\email{Clifford.F.Frez@jpl.nasa.gov}
\author{Jonathon~Hunacek}
\affiliation{Jet Propulsion Laboratory, California Institute of Technology, Pasadena, California 91109, USA}
\email{jhunacek@gmail.com}
\author{Chao-Te~Li}
\affiliation{Institute of Astronomy and Astrophysics, Academia Sinica, Taipei, Taiwan}
\email{ctli@asiaa.sinica.edu.tw}
\author[0000-0003-4063-2646]{Ian~N.~Lowe}
\affiliation{Department of Astronomy and Steward Observatory, University of Arizona, 933 N Cherry Avenue, Tucson, AZ 85721, USA}
\email{ianlowe@arizona.edu}
\author{King~Lau}\affiliation{Department of Physics, California Institute of Technology, Pasadena, California 91125, USA}
\email{kennylau@caltech.edu}
\author{Daniel~P.~Marrone}
\affiliation{Department of Astronomy and Steward Observatory, University of Arizona, 933 N Cherry Avenue, Tucson, AZ 85721, USA}
\email{dmarrone@email.arizona.edu}
\author{Evan~C.~Mayer}
\affiliation{Department of Astronomy and Steward Observatory, University of Arizona, 933 N Cherry Avenue, Tucson, AZ 85721, USA}
\email{evanmayer@arizona.edu}
\author{Guochao~Sun}
\affiliation{CIERA and Department of Physics and Astronomy, Northwestern University, 1800 Sherman Avenue, Evanston, IL 60201, USA}
\email{jsun.astro@gmail.com}
\author{Isaac~Trumper}
\affiliation{Department of Astronomy and Steward Observatory, University of Arizona, 933 N Cherry Avenue, Tucson, AZ 85721, USA}
\email{itrumper@optics.arizona.edu}
\author{Anthony~D.~Turner}
\affiliation{Jet Propulsion Laboratory, California Institute of Technology, Pasadena, California 91109, USA}
\email{anthony.d.turner@jpl.nasa.gov}
\author{Ta-Shun~Wei}
\affiliation{Institute of Astronomy and Astrophysics, Academia Sinica, Taipei, Taiwan}
\email{tashun@asiaa.sinica.edu.tw}
\author{Michael~Zemcov}
\affiliation{Rochester Institute of Technology, Rochester, NY, 14623, USA}
\email{zemcov@cfd.rit.edu}
\collaboration{27}{TIME collaboration}
\date{19 Nov 2025}

\begin{abstract}

    We present the processing of an observation of Sagittarius~A (Sgr A) with the Tomographic Ionized-carbon Mapping Experiment (TIME), part of the 2021--2022 commissioning run, to verify TIME's hyperspectral imaging capabilities for future line-intensity mapping. Using an observation of Jupiter to calibrate detector gains and pointing offsets, we process the Sgr A observation in a purpose-built pipeline that removes correlated noise through common-mode subtraction with correlation-weighted scaling, and uses map-domain principal component analysis to identify further systematic errors. The resulting frequency-resolved maps recover strong \xiicotwoone{} and \xiiicotwoone{} emission, and a continuum component whose spectral index discriminates free-free emission in the circumnuclear disk (CND) versus thermal dust emission in the 20\kms{} and 50\kms{} molecular clouds. Broadband continuum flux comparisons with the Bolocam Galactic Plane Survey (BGPS) show agreement to within $\sim$5\% in high–SNR molecular clouds in the Sgr A region. From the CO line detections, we estimate a molecular hydrogen mass of between $5.4 \times 10^5 M_\odot$ and $5.7 \times 10^5 M_\odot$, consistent with prior studies. These results demonstrate TIME’s ability to recover both continuum and spectral-line signals in complex Galactic fields, validating its readiness for upcoming extragalactic CO and [\ion{C}{2}] surveys.
\end{abstract}


\section{Introduction and Context}  \label{sec:background}
    The Tomographic Ionized-carbon Mapping Experiment (TIME) is an array of millimeter-wave spectrometers intended for line-intensity mapping (LIM) of high-redshift line emission in the singly ionized carbon [\ion{C}{2}] line and a range of rotational carbon monoxide (CO) lines~\citep{Sun_2021}. LIM experiments survey a cosmological volume in three dimensions (two spatial and one spectral), observing the aggregate emission of sources, rather than identifying specific targets~\citep{kovetz2017,Bernal_2022}.
    
    However, while CO and [\ion{C}{2}] lines act as tracers of cosmic large-scale structure in the LIM paradigm, they act locally as tracers of star formation. The primary form of star-forming gas is molecular hydrogen (H$_2$), but H$_2$ is difficult to detect directly. Instead, CO is commonly used as a tracer of molecular hydrogen~\citep{SaintongeCatinella22}. CO rotational transitions are efficient coolants for molecular clouds, with collisions with H$_2$ exciting CO molecules that radiatively decay and emit in the submillimeter regime. Hence, in resolved studies of star-forming regions, by mapping CO emission lines and their isotopologues (e.g., $^{13}$CO), one can infer molecular gas properties and conditions~\citep[see review by][]{Bolatto2013}.

    The spectral coverage of TIME (approximately 183-326 GHz) encompasses both high-redshift CO/[\ion{C}{2}] line emission and local CO line emission. To verify instrument operations and develop data reduction strategies in preparation for future LIM observations, TIME observed several galactic targets during the 2021-2022 winter observing season at the Arizona Radio Observatory (ARO) 12-meter telescope at Kitt Peak, Arizona. These sources included photo-dissociation regions, carbon stars, and dense molecular clouds, but this paper focuses on the observation and analysis of emission from and surrounding the Galactic Center radio source of Sagittarius A (Sgr A), including the Circumnuclear Disk (CND) and a small portion of the Central Molecular Zone (CMZ). 

    Sgr A is an appropriate first source of analysis, being the original source of radio astronomy with its serendipitous discovery by~\cite{Jansky33}. Subsequent high-resolution submillimeter observations have provided some revealing datasets around the Galactic nucleus. These include imaging of the supermassive black hole Sgr A$^*$ with the Event Horizon Telescope \citep{EHTSgrA} and decades of mapping of the surrounding CND and CMZ \citep[e.g.,][]{MorrisSerabyn96}. In addition to verifying TIME operations and analysis, we use our preliminary observation---a simultaneous observation of continuum, \xiicotwoone{}, and \xiiicotwoone{} emission---to illustrate physical characteristics of the molecular cloud structures in this region and to corroborate prior results. 

    The paper is organized as follows: In the rest of this section, we summarize the known properties of the Galactic Center observable at millimeter/submillimeter wavelengths and present an overview of the TIME instrument and the specific Sgr A observation under scrutiny.  In \autoref{sec:data_processing}, we detail our data calibration and reduction pipeline that transforms raw time-ordered data (TOD) into calibrated per-frequency flux maps. In \autoref{sec:spectral_analysis}, we present the resulting fit for the continuum dust emission, and validate our absolute flux calibration against an observation from the Bolocam Galactic Plane Survey (BGPS)~\citep{Ginsburg13}. We also extract integrated spectra of the CND and surrounding molecular clouds, and estimate molecular hydrogen mass from CO line-ratio analysis. Finally, in \autoref{sec:discussion}, we examine instrument-originated systematics prominent for this observation, motivate the need for improved calibration of detector spectral profiles, and discuss strategies to address these issues for future deployments.

    \subsection{Sgr A Complex in the Far-infrared Regime} \label{sec: sagA*_complex}

    \begin{figure}
        \centering
        \includegraphics[width=\linewidth]{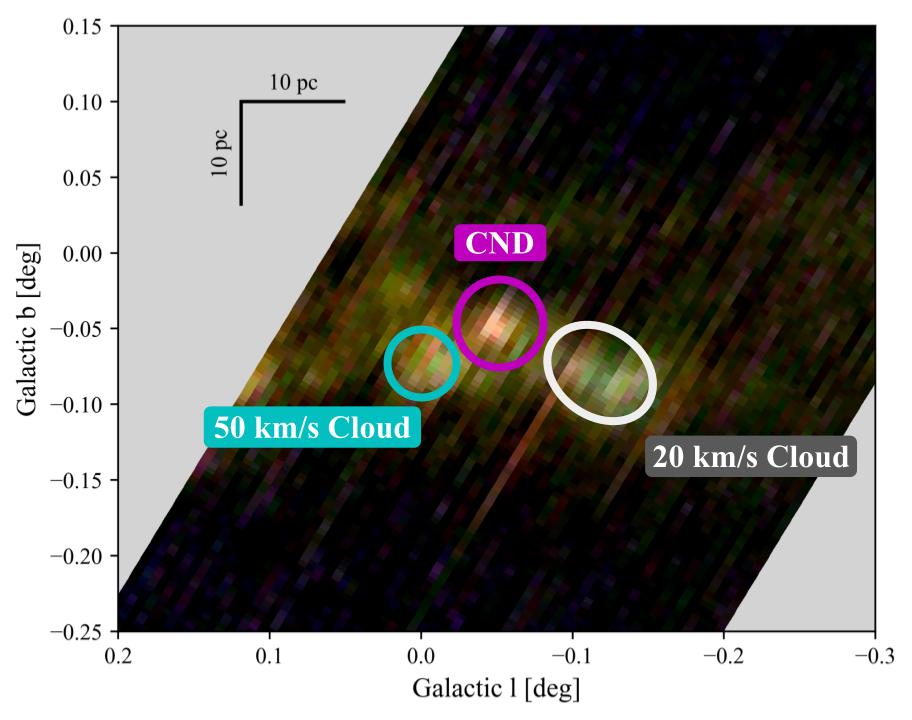}
        \caption{False–color TIME image of the Sgr A region. The frequency range 183-326 GHz spanned by TIME was linearly mapped onto the colors corresponding to 700-400 nm, and the channels were then additively blended. This image is processed with Gaussian-smoothing and a second round of PCA for pure visual demonstrative purposes.} 
        \label{fig:time_sgrA}
    \end{figure}
    
    Our analysis focuses on three regions, identified in~\autoref{fig:time_sgrA}, that are within approximately 20 pc of the Galactic nucleus. The \textbf{CND} is a dense molecular disk or toroidal structure immediately surrounding the nucleus. Enclosed within the CND is Sgr A West, a compact filamentary \ion{H}{2} region. Sgr A West itself encloses Sgr A$^*$. Here, we use the CND as a shorthand for the inner 10 pc or so of the Sgr A complex, which is unresolved due to the limited angular resolution of TIME. This includes not only the CND, but also Sgr A East, a supernova remnant that at the angular resolution of TIME is entirely confused with the CND in projection. The \textbf{50\kms{} cloud}, named for its $+50$\kms{} radial velocity, is also designated M$-0.02$$-0.07$ for its Galactic coordinates. Although observed prominently in molecular gas and dust, it is also associated with a cluster of compact \ion{H}{2} regions~\citep{Ekers83,Ho85} potentially triggered by the adjacent Sgr A East. \textbf{The 20\kms{}} cloud, likewise named for its $+20$\kms{} radial velocity, is also designated M$-0.13$$-0.08$. It too is associated with a compact \ion{H}{2} region~\citep{Ho85}, though again predominantly seen through molecular gas and dust. 
    
    The molecular clouds immediately surrounding Sgr A are some of the densest, most emissive clouds in the submillimeter regime, and the highly active nature of the CMZ enables local studies of star formation and feedback processes in galactic nuclei in comparison to high-redshift starburst environments (see~\citealt{Henshaw23} for a review). A recent innovation has been a close three-dimensional study of the CMZ and its structures through analysis of high-resolution infrared data from \emph{Herschel} (e.g.,~\citealt{3DCMZ-I,3DCMZ-II}). However, long before this, the CMZ has undergone thorough and detailed study in position-position-velocity space through CO emission in particular (including, in an extremely incomplete list of examples:~\citealt{Bania77,bally1987,Bally88,oka1998,Oka12,Tokuyama19}).

    Observing the Sgr A complex with TIME allows us to map CO line emission and continuum dust emission across the region, and also to cross-check our broadband continuum and spectral analyses against the considerable preceding body of work on the CMZ. This very local map of line intensity is thus a crucial step towards the extragalactic LIM we aim to do with TIME.

    \subsection{TIME Instrument Overview} \label{sec:TIME_Mechanical_overview}

        The TIME instrument is a multi-staged, milli-Kelvin cryostat with the coldest stage operating at a temperature of $\sim250$ mK. In this stage are held two (two designed, one online for this engineering run) $R \sim 100$ curved, diffraction grating spectrometers that are machined out of aluminum and gold-plated. Each spectrometer samples one polarization angle through rectangular waveguides, which also act as low-pass filters. These are coupled to a linear array of 16 feedhorns per spectrometer, which comprise the $1.3^\circ\times0.45'$ TIME field of view on the sky.  
        
        Mounted on the back of each spectrometer are six detector modules, comprising sixty spectral channels (f) and sixteen spatial channels (x). The electrical readout configuration is multiplexed to reduce wiring and improve noise/thermal performance using a time domain multiplexed (TDM) system~\citep{Batt2008}. Thus, the $16\times60$ detectors are grouped into 33 rows (multiplexer row index or mux-r) and 32 columns (multiplexer column index or mux-c), where each column is electrically biased together. Both coordinate conventions will be used in this analysis. The detectors are transition-edge sensor (TES) bolometers with an absorber made of gold and silicon-nitride coupled to aluminum and titanium TES. The detectors are fabricated in subarrays, and 24 subarrays cover the focal plane behind each spectrometer. Two different absorber designs have been created to optimize performance in the low and high limits of TIME's bandwidth, approximately 183-326 GHz. While this bandwidth was chosen for high-redshift cosmological surveys and covers $z\gtrsim1$ CO and [\ion{C}{2}] emission, the instrument's spectral resolution and wide coverage are valuable for Galactic science as it covers local emission in the \xiicotwoone{} line and many other common dense-gas tracers.

        \cite{Crites_14} and \cite{butler2024} provide additional details of the TIME instrument design and implementation, although we have described above the details most relevant to the present work.

    \subsection{Observation and Raw Data} \label{sec:status}

        On February 8, 2022, we conducted a two-dimensional raster scan of Sgr A complex using the ARO 12-meter telescope on Kitt Peak, Arizona, as part of an engineering campaign. This dataset offers both a strong astrophysical signal with notable spectral structure and substantial noise due to the low observing elevation and instrumental noise. This special combination makes it a good testing ground for data cleaning and map-making strategies.

        \begin{figure}[hbt]
            \centering
            \includegraphics[width=\linewidth]{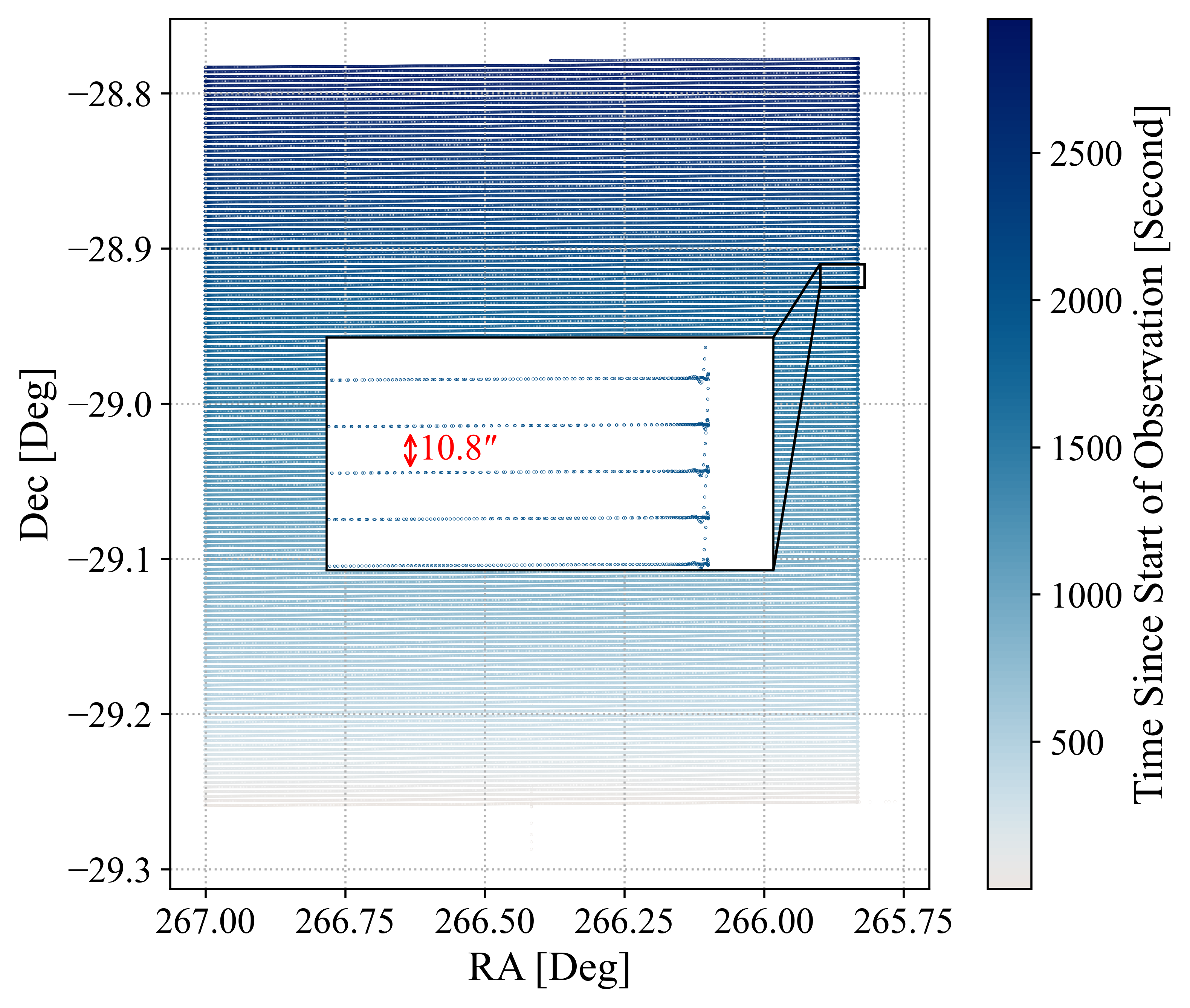}
            \caption{Scan pattern for the observation of Sgr A on February 8, 2022. The telescope steps in declination after completing each sweep in right ascension, producing 2D coverage of the field over the observation duration.}
            \label{fig:scanpattern}
        \end{figure}

        The telescope performed the raster scan by tracking a fixed declination (Dec), scanning across right ascension (RA), and then stepping to track the next declination in the raster pattern, as shown in~\autoref{fig:scanpattern}. 
        Consequently, all intermediate maps presented in the data processing sections are shown in RA/dec coordinates, made from TODs that are transformed to the J2000 frame from the apparent frame as appropriate. However, all final science products are manually transformed into Galactic coordinates to align with the conventional reference frame for the Galactic Center and Sgr A.
        
        TIME also observed Jupiter as a calibrator in a similar 2D raster scan, approximately 2.5 hours after the Sgr A observation. We outline the basic parameters of both the Sgr A and Jupiter observations in~\autoref{tab:obs}.

        \begin{deluxetable*}{cccccccccc}
            \tablecaption{TIME observations on Feb 8, 2022 used for this work.\label{tab:obs}}
            
            \tablehead{
            \colhead{Target} & \colhead{Local time} & \colhead{RA} & \colhead{Declination} & \colhead{Frame} & \colhead{$\tau_{0,225\text{GHz}}$} & \colhead{Elevation} & \colhead{Raster size} & \colhead{Raster step} & \colhead{Scan speed}\\
            \colhead{} & \colhead{(MST)} & \colhead{(hr:m:s)} & \colhead{(deg:m:s)} & \colhead{} & \colhead{} & \colhead{(deg)} & \colhead{(deg $\times$ deg)} & \colhead{(deg)} & \colhead{(deg s$^{-1}$)}
            }
            \startdata
            Sgr A&08:36&17:45:40&-29:00:28.8&J2000.0&0.12&28&$1.0\times0.5$&$0.003$&$0.125$\\
            Jupiter&10:50&22:43:55.7&-9:05:21.5&Apparent&0.11&29&$0.3\times0.5$&$0.005$&$0.05$
            \enddata
            \end{deluxetable*}
        
\section{Data Processing}\label{sec:data_processing}

\begin{figure*}[t]
        \centering
        \includegraphics[width=\linewidth]{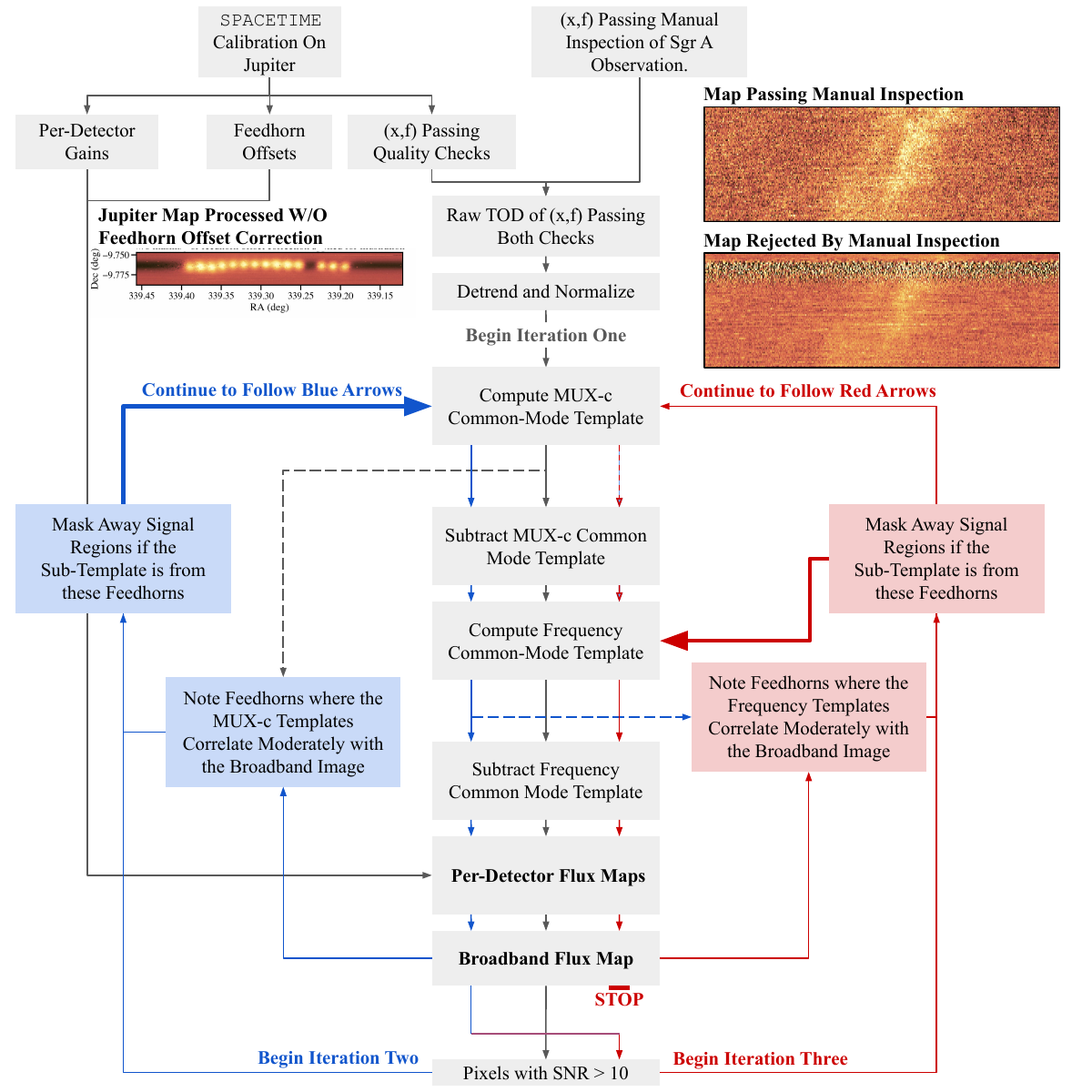}
        \caption{TOD Processing Pipeline: The workflow begins by using \textit{SPACETIME} to determine per-detector gains (~\autoref{sec:gains}) and feedhorn offsets (\autoref{sec:feedoffset}). Raw TODs passing quality checks are filtered, normalized (\autoref{sec:highpass}), and corrected using mux-c and frequency common-mode templates (\autoref{sec:common-mode-noise}). Per-detector calibrated flux maps are generated (\autoref{sec: processing_map}).To read this flowchart, follow the gray arrows for the general preparation of data and the first iteration of common-mode noise removal, continue along the blue arrows for the second iteration, and proceed to the red arrows for the third iteration. Each arrow is traced once. Detector mapping conventions such as (x,f) and mux-c are defined in~\autoref{sec:TIME_Mechanical_overview}}
        \label{fig:processing_flowchart}
    \end{figure*}

    The transformation from raw detector TOD to science-ready maps involves a sequence of calibration, cleaning, and projection steps designed to preserve astrophysical signal while mitigating correlated readout and systematic noise. \autoref{sec:calibration_with_spacetime} covers the calibration procedure, which yields a frequency-dependent detector count-to-flux conversion factor.  In \autoref{sec:processing_timestream}, we describe the processing of Sgr A TODs. In \autoref{sec: processing_map}, we discuss how these TODs are folded into flux maps and further processed. We also summarize the TOD processing workflow in~\autoref{fig:processing_flowchart}. 

    \subsection{Calibration using \textit{SPACETIME}} \label{sec:calibration_with_spacetime}
    
        We perform calibration using \textit{SPACETIME} (Sky Pointing and Amplitude Calibration Estimator for TIME), a data-processing pipeline that we are developing to be generalizable across all observations (paper in preparation). Here, we describe the calibration procedures implemented in the version of \textit{SPACETIME V1.0} current as of the present work. 
    
        \subsubsection{Low-level Data Filtering During Calibration}
            \label{sec:low_level_data_filtering_cal}
            To convert raw detector readouts into flux measurements across frequency channels, we calibrate the data by comparing the response to the observation of a well-understood source, in this case, Jupiter observed 2.5 hours after Sgr A. The planet-based calibration yields a flux conversion factor---hereafter referred to as the gain---for each detector (i.e., at each frequency and for each feedhorn). We may then apply these gains to the Sgr A observation to enable further processing.
            
            We process the planet data in \textit{SPACETIME} as follows. First, the time-ordered pointing and flux data are spliced into separate scans in RA. In each of these scans, we then mask the planet---which is bright enough to be easily identified in the TOD---with a generic peak finding algorithm and filter with a fifth-order polynomial. We may then construct a 2D map of Jupiter for each detector by binning the filtered TOD into a grid in RA/dec. To each map we fit a Gaussian beam, parameterized by two center coordinates $C_x$ and $C_y$, two FWHM $\sigma_x$ and $\sigma_y$, and an overall amplitude $A_c$. The center coordinates serve to estimate the location of Jupiter. These fits are further used to exclude detectors with poor fit quality. With this na\"{i}ve fit, we generate a mask encompassing the extent of Jupiter in the map.
            
            We then revert back to the TOD and repeat the process, but now using the newly generated mask to exclude TOD samples falling on Jupiter from the polynomial fit, so as to reduce signal loss when filtering. In addition, a per-scan variance is estimated based on the masked residuals. At this time, we also correct for the atmospheric flux attenuation in TOD space,
            \begin{equation}
                C_{\text{T}} = C_{\text{O}} \times {e^{\tau_0 \times A}}.
            \end{equation}
            Here $C_i$ denotes ``counts", which are unitless detector outputs with an arbitrary scale that can vary significantly from detector to detector.  Specifically, $C_{\text{T}}$ is the true counts value we would have measured from Jupiter without atmospheric attenuation, while $C_{\text{O}}$ is the counts value we observed in TOD space, and $\tau_0$ is the frequency-dependent atmospheric optical depth at zenith. We use the am model developed by \citet{paine2019atmospheric} to obtain a template in $\tau_0$ at a reference frequency of 225 GHz. We then use the same family of models consistently to extrapolate the frequency dependence of $\tau_0$.  This still requires modification by $A$, the air mass at the zenith angle of the telescope at the time each data point was recorded. Uncalibrated maps are regenerated using the same map-mapping approach as the na\"{i}ve maps, but now each cell is inverse-variance weighted according to the per-scan variance. Finally, a Gaussian beam is re-fit to each of these uncalibrated maps, and planet maps with poor fits or low signal-to-noise are removed. For the remaining planet observations, the best-fit amplitude is used to calculate the gain, as described in more detail in~\autoref{sec:gains}. 
    
        \subsubsection{Feedhorn Offsets Correction} \label{sec:feedoffset}
            The linear array of 16 TIME feedhorns map to an approximately linear array of positions on the sky. As the telescope performs a raster observation with the array aligned to the scan direction, the feedhorns cross the same point on the sky at different times. This can be seen in~\autoref{fig:processing_flowchart}, where we do not correct the offsets between feedhorns and therefore reconstruct a map with a linear array of Jupiter images along the declination direction, reflecting the true feed offsets. The missing beam was a dark feedhorn during that observation. We must correct for this by estimating the change in RA and dec per feed with reference to a reference position, which we take as feedhorn 8. Using our fits on the Jupiter observation described in the previous section, we can estimate a consistent point in space via the fit parameters, $C_x$ and $C_y$. We then take the mean difference between each planet center and the reference feed's planet center to calculate the relative pointing. This is later used to correct for the relative offsets in the Sgr A dataset.  
    
        \subsubsection{Obtaining Per-Detector Gains}
        \label{sec:gains}
            The final step for estimating gains is to produce a fiducial model of the planet's flux. This is done through a Planck function $ B(\nu; T_b(\nu))\,$ wherein we use frequency-dependent brightness temperatures, $T_b$, per the model reported from CASA by \citet{butler2012flux} and \citet{bean2022casa}, e.g, to account for the deviations of Jupiter's spectrum from that of a perfect blackbody. 
        
            \begin{equation}
                I_{\nu} = B(\nu; T_b(\nu))\,\Omega_p\ 
            \end{equation}
            The angular size of the planet, $\Omega_p$, computed via Astropy's Solar System Ephemeris implemented based on the JPL ephemeris models, is used to convert to Jansky (Jy)~\citep{park2021jpl,robitaille2013astropy,price2018astropy,price2022astropy}. 
            
\begin{figure*}
                \centering
                \includegraphics[width=\linewidth]{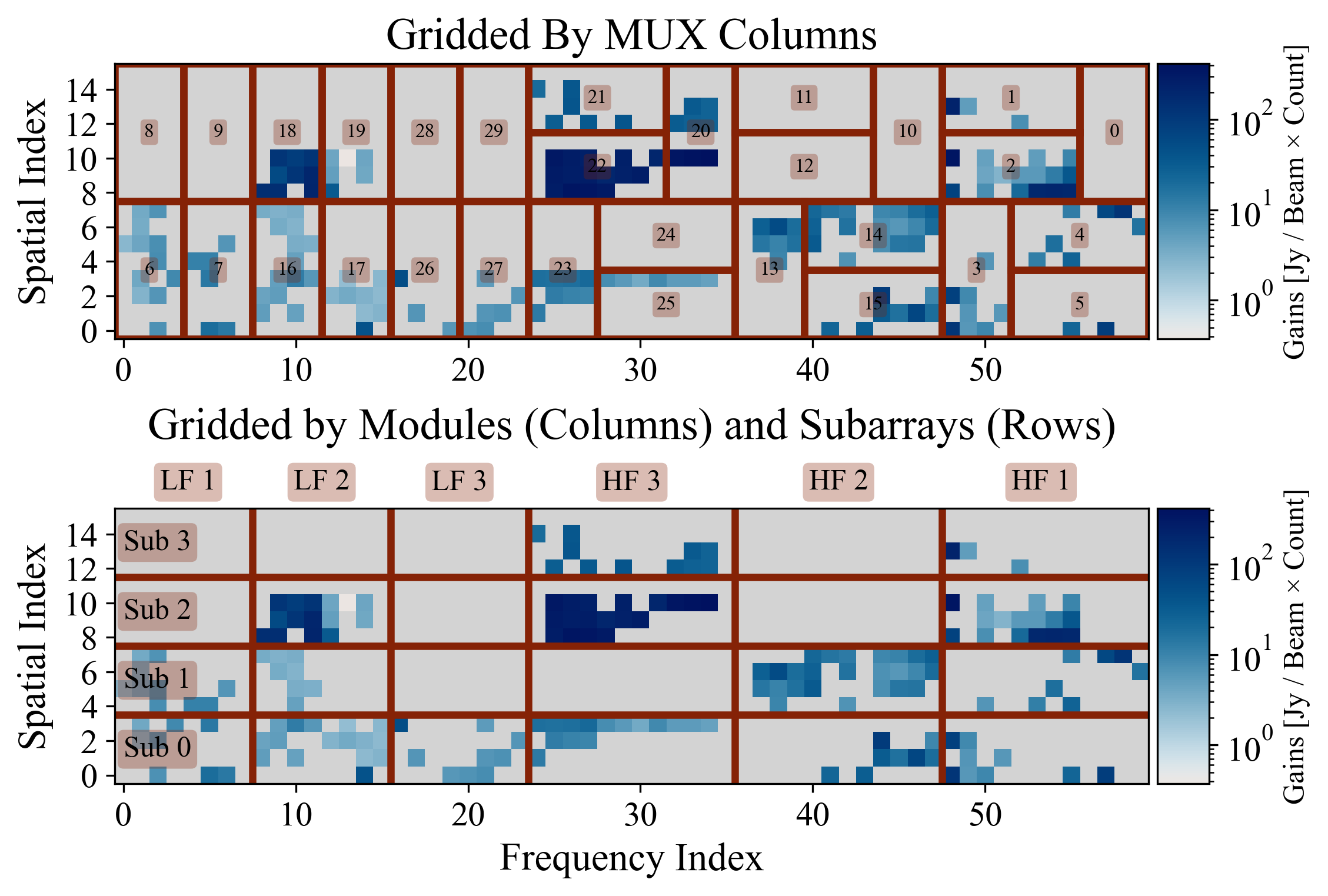}
                \caption{Per–detector gain factors derived from Jupiter observations on February 8, 2022. The gain is the multiplicative factor converting raw detector counts to Jy, as described in~\autoref{sec:gains}. Gaps correspond to detectors that did not pass quality checks.}
                \label{fig:gains_per_det}
            \end{figure*}

            The measured beam amplitude, $A_c$, in counts is obtained by fitting a 2D Gaussian to each detector’s map. The ratio of the fiducial model to $A_c$ gives a gain $G(x,f) = \frac{I_{\nu}(f)}{A_c(x,f)}$, but note that this is operating on the attenuation corrected TODs as described in~\autoref{sec:low_level_data_filtering_cal}. We show this estimated gain for each detector that passes all aforementioned quality checks in~\autoref{fig:gains_per_det}. Subarray two stands out with about an order of magnitude larger gain than the others, indicating that these detectors respond more efficiently to incoming power.
            
    \subsection{TOD Filtering} \label{sec:processing_timestream}

    We manually inspected Sgr A detector maps and removed those that have excessive noise and significant crosstalk or negative shadows \footnote{We made the choice of manual inspection because Sgr A has a less defined geometry across frequencies and a more prominent diffusive region. In future iterations of the workflow, we will be able to make this quality cut on a more quantifiable basis for suitable targets.}. For example, \autoref{fig:processing_flowchart} shows an example of a map that passes manual inspection alongside one that does not. We find 179 detectors that pass both the quality cuts during the calibration process and this manual inspection, and only use readouts from those detectors for the rest of the analysis. 

    The TOD filtering takes place in two major steps: a high-pass filter detailed in~\autoref{sec:highpass}, and a common-mode fluctuation template identification and subtraction pipeline detailed in~\autoref{sec:common-mode-noise}.

    \subsubsection{Basic High-pass Filtering and Normalization}
    \label{sec:highpass}

    We removed long-time-scale drifts in the detector TOD using a Fourier-domain low-pass filter. The filter had a Gaussian roll-off with $\sigma = 0.02$ Hz at the cutoff edge frequency of 0.01 Hz. Prior to filtering, a linear trend was subtracted from each detector’s TOD to remove first-order drifts. Next, we estimate the fluctuation amplitude of each detector by computing the standard deviation of the remaining TOD, restricted to the central $50\%$ of the time samples to avoid edge effects. This quantity serves as a normalization factor. Finally, we normalize the cleaned TOD by dividing by this estimated amplitude, resulting in approximately zero-mean TOD with standardized variance. 

    \subsubsection{Common-mode Noise Removal} \label{sec:common-mode-noise}

\begin{figure*}
        \centering
        \begin{subfigure}{\linewidth}
            \centering
            \includegraphics[width=\linewidth]{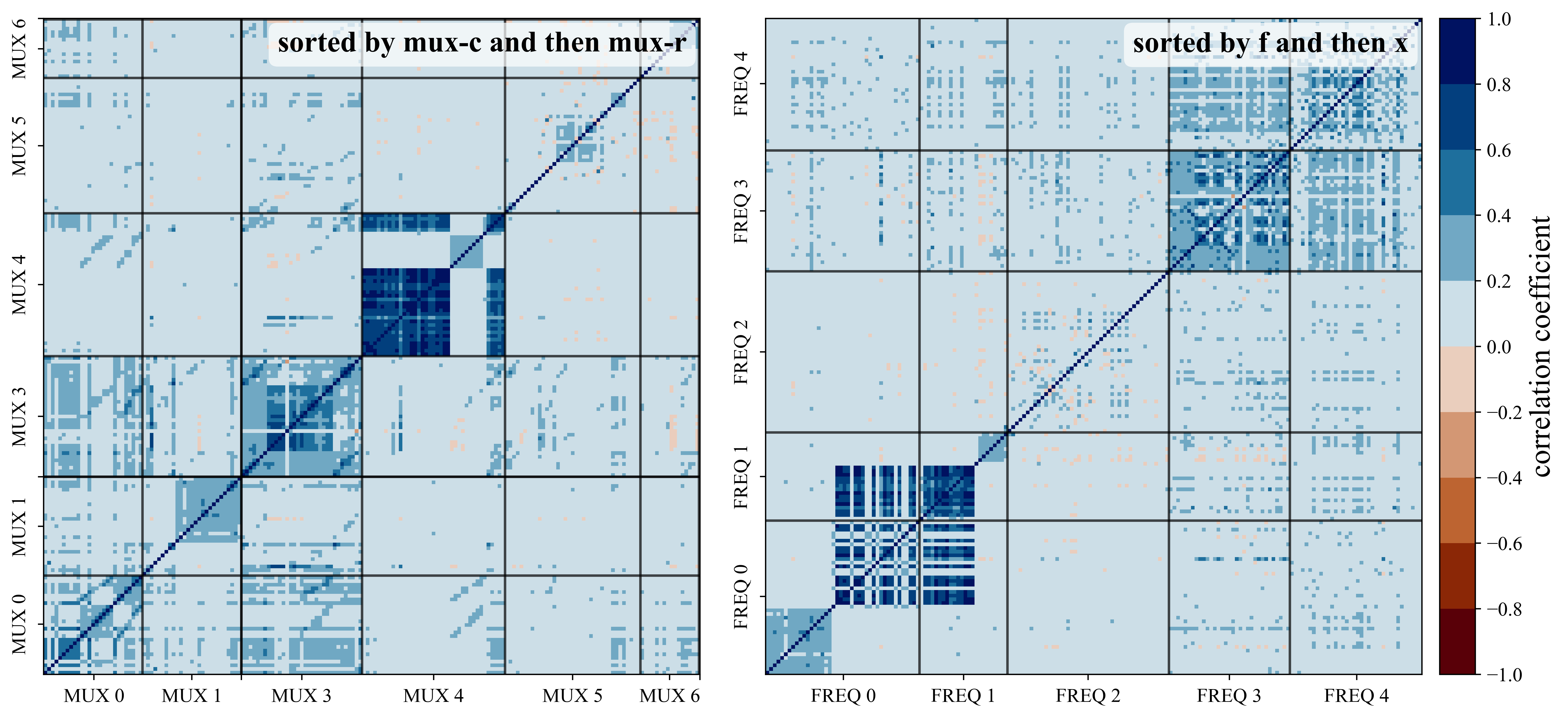}
            \label{fig:correlation_before_tempsub}
        \end{subfigure}

        \begin{subfigure}{\linewidth}
            \centering
            \includegraphics[width=\linewidth]{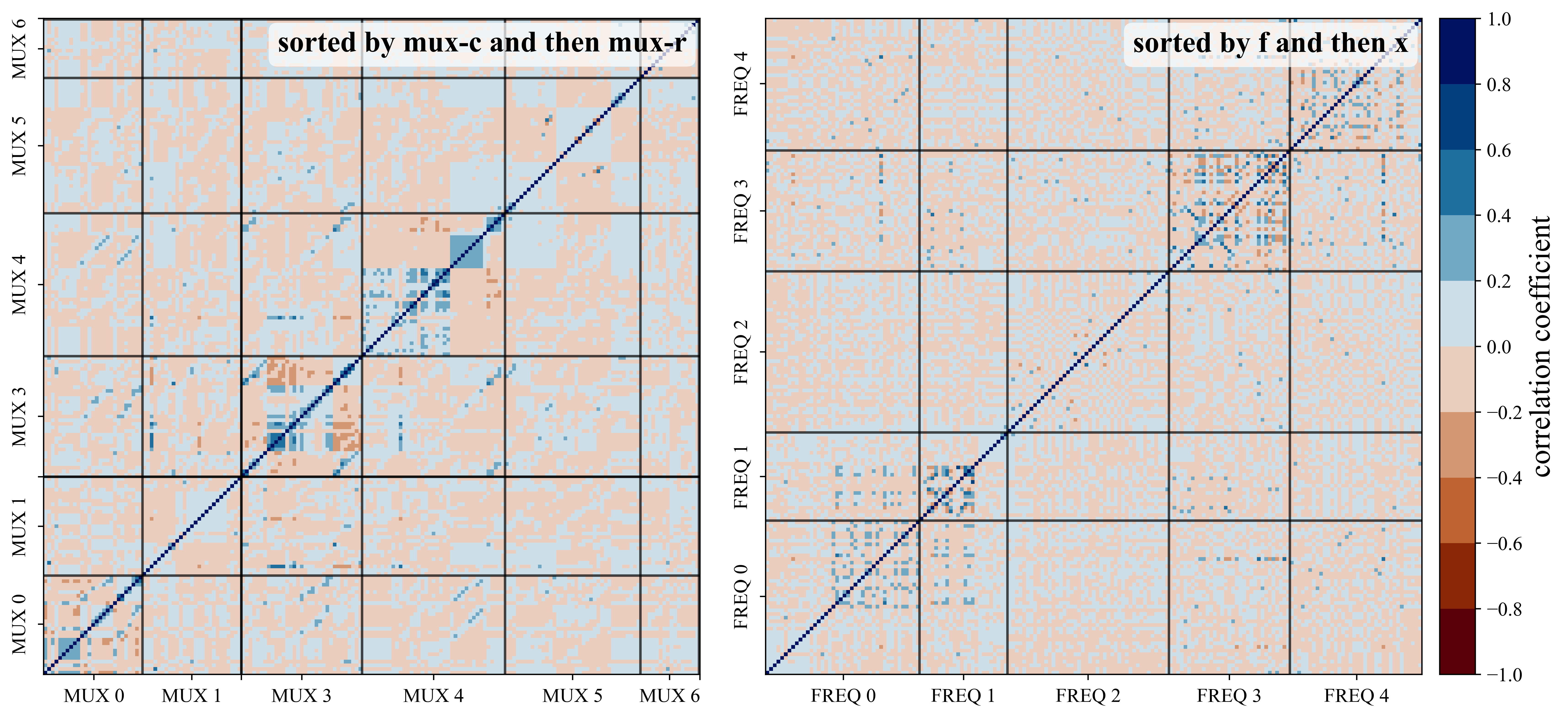}
            \label{fig:correlation_after_tempsub}
        \end{subfigure}

        \caption{Top: Correlation matrices of normalized detector TOD before common-mode template subtraction. The left panel shows the correlation coefficients for all 179 detectors, ordered by mux-c and then mux-r, with grid lines indicating mux-c-block templates. The right panel shows the same data ordered by frequency and spatial index, gridded by frequency-block templates. Bottom: Same as above, but for the correlation matrix after common-mode template subtraction. Blocks of high correlation indicate strong common-mode noise components, likely of instrumental origin (see~\autoref{sec:discussion}).}
        
        \label{fig:correlation_of_detector}
    \end{figure*}

    We expect significant correlated noise from atmospheric fluctuations, and we also noticed correlated noise in detector subarrays and readout electronics. To assess the dominant structure of this correlation, we computed the pairwise correlation coefficients between all 179 functioning detectors, organizing the results by the observing multiplexer column (mux-c) or frequency, as shown in~\autoref{fig:correlation_of_detector}. 

    To partially remove this correlated noise after initial pre-processing, we implement a two-stage, correlation-weighted common-mode template subtraction procedure. We now describe the subtraction procedure and elaborate on template construction later in this subsection. The principle of the common-mode calculation is such: The signal occurs at different regions of the map for detectors in each feedhorn and has different spatial structure. Therefore, noise shared in detector maps dominates the common-mode template However, due to the limited number of detectors that passed the aforementioned quality checks, we had to construct common-mode templates from groups of four mux-c and groups of twelve frequency indices as opposed to finer grouping, such as by mux-c or by subarray. We first compute the mux-c-based common-mode templates by taking the median TOD across all detectors falling into each block. For each detector, we evaluated the correlation coefficient $r$ between its normalized TOD and the corresponding mux-c block template. 
    
    If the common-mode template correlates completely with the detector TOD, we wish to subtract the template completely; if it anti-correlates completely we wish to subtract its negative; if it has no correlation we wish to do nothing with the template, since we would risk adding time-correlated noise by applying the template in any way. Since the detector--template correlation coefficient $r$ will never be exactly zero or $\pm1$, we smoothly interpolate between these extremes with a sigmoid function calculated from $r$,
    \begin{equation}
        \mu = 
        \left[
            \frac{1+\operatorname{erf}(20|r| - 6)}{2}
        \right]^{\eta},
    \end{equation}
    so that the template is scaled by $\pm\mu$ depending on the sign of $r$ and then subtracted from the detector TOD. Note also that the exponent $\eta$ depends on the detector’s frequency index.
    To suppress contamination from strong spectral features, we flag channels
    corresponding to known spectral lines (e.g., \xiiico, $^{12}$CO, and HCN) and exclude them from the calculation of common-mode templates. As a result of both this and the presence of strong line signals, we expect intrinsically lower correlation between the common-mode template and the detector TOD in those channels. Therefore, to boost the scaling magnitude $\mu$ for these flagged channels to ensure sufficient removal of noise, we set $\eta = 1/4$; for all others, we set $\eta = 1$. The boost is fairly subtle: for $\eta=1$, $\mu(r=0.3)=0.5$; for $\eta=1/4$, $\mu(r=0.25)=0.5$.

    After subtracting the scaled template, we compute the standard deviation of each TOD, restricted to the central $50\%$ of samples to avoid anomalies towards the start and end of the observation, and use it to normalize the signals once more. From these mux-c block subtracted TOD, we carry out the same procedure described above for frequency blocks. Finally, we restore physical units by reapplying the previously computed normalization factors. 

    We follow an iterative procedure to construct these mux-c blocks and frequency block templates so that they do not excessively correlate with the astrophysical signal (represented by the gray, blue, and red paths in~\autoref{fig:processing_flowchart}). This step was necessary to systematically identify and apply special handling to TOD containing exceptionally strong astrophysical signals before they go into template construction. We assume that templates containing strong signals would correlate more strongly with the broadband continuum sky signal than templates dominated by common-mode noise. Furthermore, we expected those strong signals to occur predominantly in pixels with a high broadband signal-to-noise ratio (SNR). Based on these assumptions, we iterated the template subtraction process three times.
    
    \paragraph{First iteration (gray path in~\autoref{fig:processing_flowchart})} We construct mux-c-block and frequency-block templates na\"{i}vely as previously described. Using these, we generate a na\"{i}ve set of per-detector uncalibrated maps and follow the procedures outlined in~\autoref{sec: processing_map} to produce an initial broadband uncalibrated map and corresponding error map. We identify high-SNR pixels by selecting those with $\text{SNR}>10$. 
    
    \paragraph{Second iteration (blue path in~\autoref{fig:processing_flowchart})}Because templates are assembled from TOD associated with detectors across the focal plane, we correct for feedhorn-dependent spatial offsets and compute the correlation between each offset-adjusted template and the broadband continuum image. We flag any feedhorn producing a correlation coefficient $r>0.35$ for each template\footnote{This threshold and the threshold for the third iteration were selected empirically during the early stages of the analysis and subsequently held fixed for consistency across later revisions. While this choice was guided by practical considerations rather than a formal optimization, we found it sufficient for the purposes of the present study.}. While reconstructing the mux-c-block templates for this iteration, we mask around the high SNR regions in TOD space whenever the associated TOD originated from a flagged feedhorn. This prevent the strong astrophysical signals from contributing to the updated mux-c-block template. However, the new frequency-block templates absorbed some of the structure resembling the broadband signal, and we will address this in the third iteration. We now compute an interim broadband continuum map and error map, and identify a new set of high-SNR pixels. 
    
    \paragraph{Third iteration (red path in~\autoref{fig:processing_flowchart})}We repeat the strong astrophysical signal structure identification procedure for the frequency-block templates from the second iteration with a correlation coefficient $r>0.25$. We apply the same masking procedure for these frequency templates. After completing all three iterations, the mux-c-block templates showed substantially reduced signal leakage compared to the initial, na\"{i}ve templates, as shown in~\autoref{appendix:cm-template-supplement}, where we fold these templates into maps for visualization. While the frequency-block templates retained residual features, we did not perform additional iterations, as their subtraction is much less aggressive than that of the mux-c-block templates. Using this final set of refined templates, we proceeded with the remainder of the template subtraction pipeline as described above. 
    
\begin{figure}
        \centering
        \includegraphics[width=\linewidth]{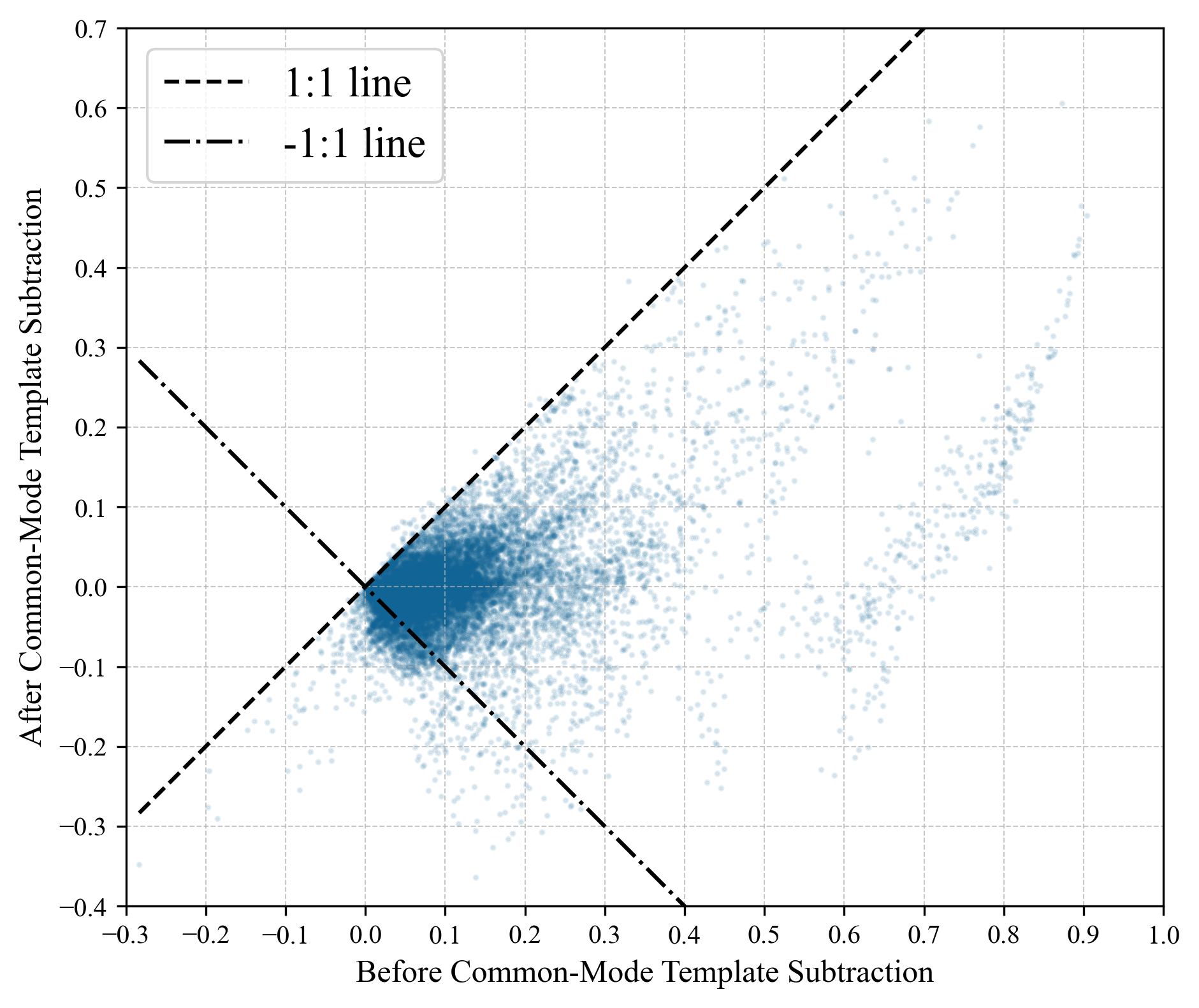}
        \caption{Detector-pair correlation coefficients before common-mode template scattered with detector-pair correlation coefficients after common-mode template subtraction. Almost all points fall below the 1:1 line, indicating an overall reduction in correlation after template subtraction. Most points also fall above the 1:-1 line, confirming that most detector pairs did not become anti-correlated at a stronger level. }
        \label{fig:scatter_template_r}
    \end{figure}

    We once again compute the pairwise correlation coefficients between the detectors to confirm the effectiveness of this common-mode template subtraction on correlated noise. The high correlation blocks previously observed have been reduced, shown in~\autoref{fig:correlation_of_detector}. While we are able to reduce the correlated structures compared to their pre-common-mode template subtraction levels, there remain residual patterns in both mux-c-block space and frequency-block space. These features are weaker but still somewhat spatially coherent, suggesting that not all instrument-origin correlations are fully removed by our current template construction and subtraction scheme. We also show a scatter plot comparing the detector-pair-wise correlation coefficient before and after this common-mode template subtraction in~\autoref{fig:scatter_template_r}. Out of the $15931$ unique detector pairs, $14240$ are below the 1:1 line and above the 1:-1 line, indicating that the common-mode template subtraction suppresses inter-detector correlations. 
    We discuss this correlated noise of apparent instrumental origin in~\autoref{sec:discussion}. The present approach already accomplishes substantial noise cleaning, and we defer more aggressive treatments to future iterations of the data-processing workflow. 

    \subsection{Sgr A Map Processing} \label{sec: processing_map}

    Having completed calibration and some data-cleaning operations on the per-detector TOD, we now fold them into flux maps and transition to data-processing in the map domain. We first correct for the atmospheric attenuation using the same method consistent with our correction for Jupiter in~\autoref{sec:low_level_data_filtering_cal}, but with parameters for this Sgr A observation in~\autoref{tab:obs}. We generate 2D raster maps with a pixel size of $14.4''$
    in RA and $21.6''$in Dec. We also correct for its feed offset computed from the Jupiter observation. We then calibrate these detector maps by the respective gains computed in~\autoref{sec:gains} and propagate gain errors accordingly. 

\begin{figure}
        \centering
        \includegraphics[width=\linewidth]{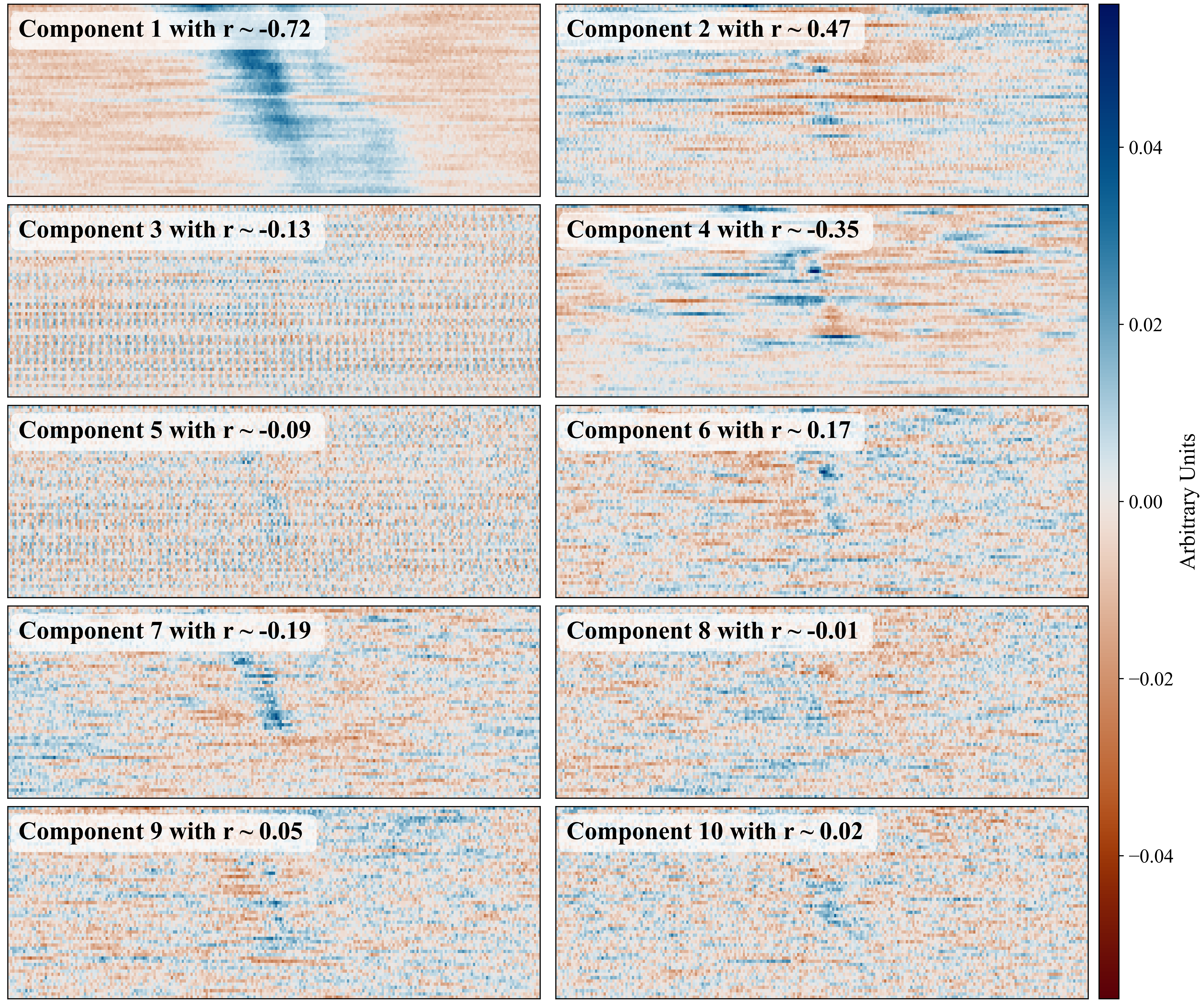}
        \caption{The ten principal components (PCs) from PCA decomposition of the per–detector maps. PCs three, five, eight, nine, and ten have correlation coefficients with the broadband continuum map below $0.15$ and are classified as systematics, which are subtracted from the data.} 
        \label{fig:10components}
    \end{figure}

    To identify any remaining systematics, we apply principal component analysis (PCA) on our per-detector flux maps. To avoid overfitting to any one detector map, we need to normalize each map before running the PCA. This normalization is chosen to be the standard deviation of the off-source region of the map. To isolate the off-source region, we mask out a parallelogram with a base of $0.4^{\circ}$ degrees in RA, height of $0.366^{\circ}$ in Dec, a skew of $0.2^{\circ}$, and centered at (RA, Dec) = (266.37$^{\circ}$, -29.057$^{\circ}$). Using Scikit-learn’s implementation of PCA \citep{pedregosa2011scikit}, we decomposed the normalized maps into ten orthogonal components as shown in~\autoref{fig:10components}. Each principal component represents a spatial pattern that captures variance common across frequencies. These components may be treated as candidate systematic modes when their correlation with the broadband continuum signal is weak.  With a correlation coefficient cut-off\footnote{This threshold was determined by inspecting the components at various cutoff values during earlier stages of the analysis.} of 0.15, we identify components three, five, eight, nine, and ten as components dominated by systematics. Components three and five are primarily composed of rapid oscillations, which are only weakly correlated with the aliasing introduced by our binning scheme. Components eight, nine, and ten show structures that superficially resemble the astronomical signal; however, given their low correlation with the na\"{i}ve broadband continuum flux map, we interpret these features as shadow residuals arising from the common mode template subtraction process, as discussed in~\autoref{appendix:cm-template-supplement}. 
    We calculate the component amplitudes by projecting each PCA component onto the original maps, and we scale these components by this projection amplitude before subtracting them from our per-detector flux maps. We can now estimate the uncertainties on our per-detector flux map, which will be necessary for subsequent analysis. 

    To estimate the error per scan for these flux maps, we proceed in two steps. We once again want to separate regions where instrumental noise is the dominant contribution, rather than the astrophysical signal. For this purpose, we apply the same parallelogram mask described above. We now evaluated the standard deviation of the masked map along the scan direction (RA) as an estimate of the per-scan noise. 
    
    Using these error maps as weights to the per-detector flux maps, we compute the inverse variance weighted average of all maps at a common frequency, which we refer to as the per-frequency flux maps. In fitting to the Jupiter observation, we found that the effective beam FWHM is about 33’’-35’’ across detectors in our science frequency band.  
    
    \begin{figure*}
        \centering
        \includegraphics[width=\textwidth]{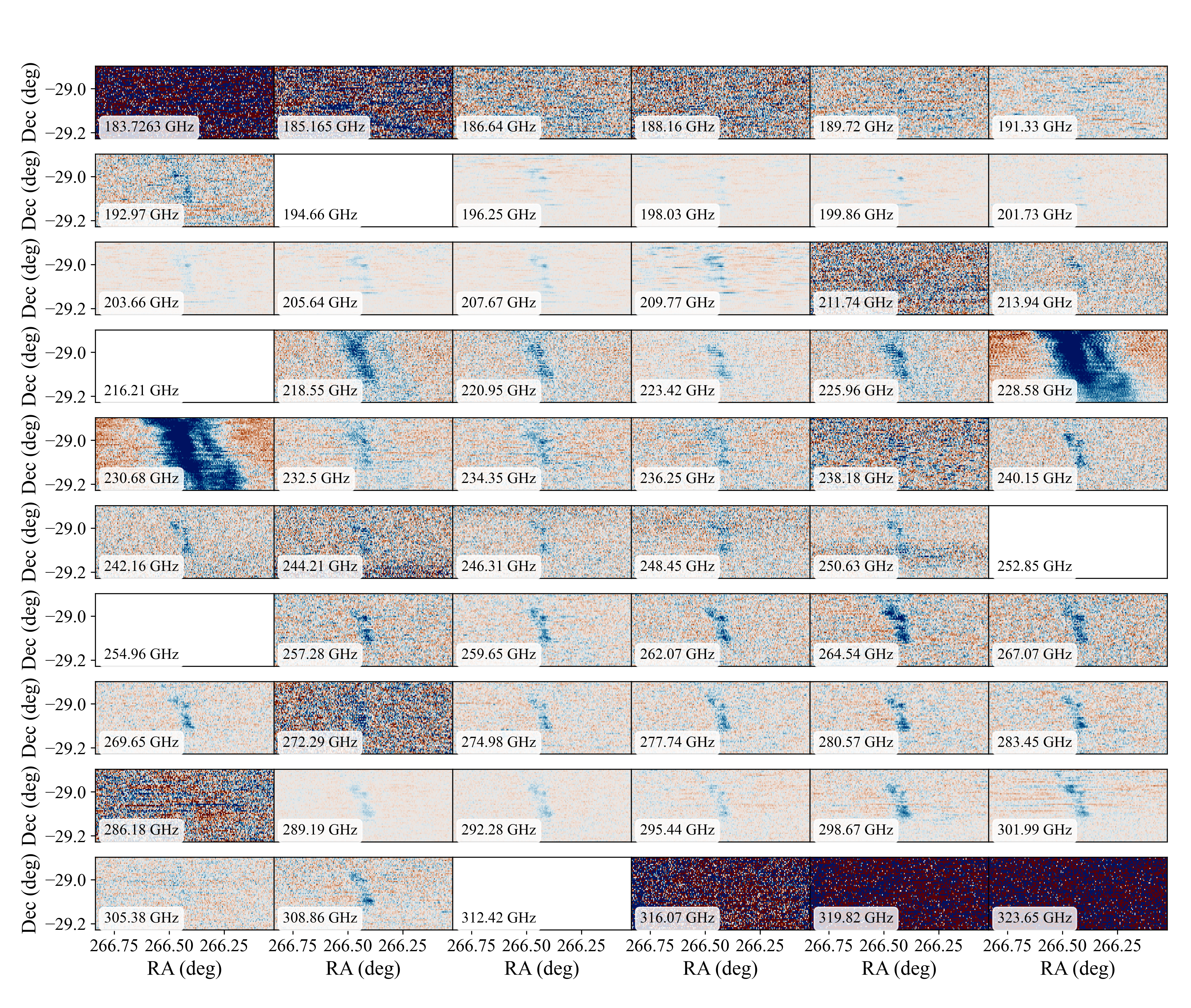}
        \caption{TIME per–frequency flux maps of the Sgr A region show on a common flux scale spanning $(-10,10)$ Jy beam$^{-1}$. Strong $^{12}$CO (2–1) emission appears at $\approx230$ GHz and diffused $^{13}$CO (2–1) emission at $\approx220$ GHz, while other channels trace a mix of thermal dust continuum and weaker molecular lines. The lowest and highest frequency channels are noise-dominated because the signal is strongly attenuated by the atmosphere.}
        \label{fig:sgr_spec}
     \end{figure*}

\section{Analysis} \label{sec:spectral_analysis}

    We analyze the per-frequency flux maps to extract spectral properties of the Sgr~A region, and we analyze the broadband continuum maps to validate our absolute flux calibration. Our workflow begins by fitting the continuum spectral index to verify the internal flux calibration across the frequency range in~\autoref{sec:spectral_index}. We then validate the absolute calibration through a broadband continuum flux comparison with the Bolocam Galactic Plane Survey (BGPS) in~\autoref{sec:broadband-flux-map}. Finally, we extract integrated spectra from key astrophysical regions in~\autoref{sec: integrated_spectra} and perform $^{12}$CO/$^{13}$CO line–ratio analysis to estimate molecular gas masses in~\autoref{sec:co_line_analysis}. The following subsections provide a detailed description of each analysis. 

    To define the 50\kms{}, CND, and 20\kms{} regions used in our analysis, we construct high SNR contours from broadband continuum maps made from an inverse-variance weighted average of our per-frequency flux maps. First, we generate an initial high-SNR binary mask for pixels with SNR greater than 8.5. We then smooth this mask slightly using a small 2D Gaussian kernel with sub-pixel widths to suppress pixel-scale noise fluctuations. After convolution, all nonzero pixels are considered to be high-SNR. These contours serve as the basis for defining the spatial boundaries of key structures. In practice, we manually divide the high-SNR contour into subcontours corresponding to each region. This approach is our best effort to describe the regions of interest, given the limited resolution of these sources.

    \subsection{Spectral Index of the Continuum } \label{sec:spectral_index}

    To characterize the spectral behavior of the continuum, we fit a power-law spectrum ($S_\nu = b\nu^{\alpha}$) to each pixel in the per-frequency flux maps. Astronomical signals in channels below 196 GHz and above 300 GHz are strongly attenuated by the atmosphere, and we exclude them from the following analysis. We also masked specific channels (5) associated with known molecular emission lines to prevent line contamination in the continuum spectral index analysis. 

    \begin{figure*}
        \centering
        \includegraphics[width=\linewidth]{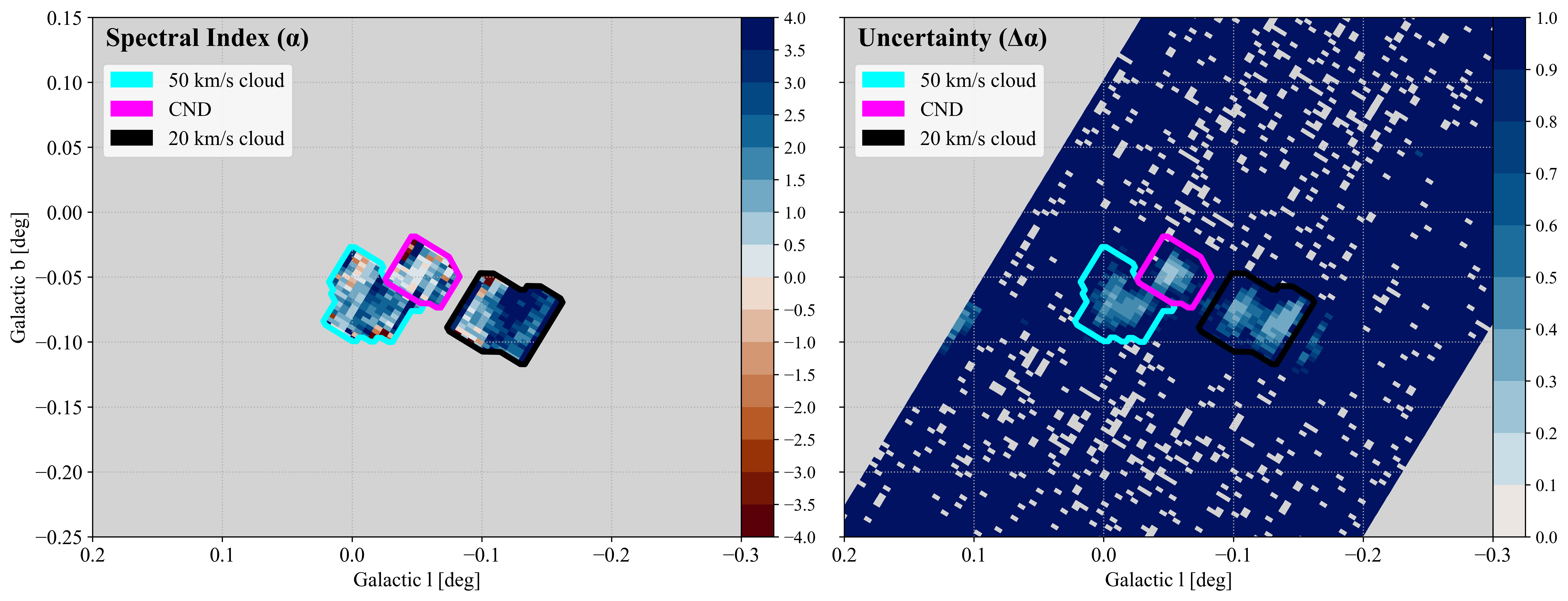}
        \caption{Left: Spectral index map derived from continuum–only frequency channels between 196.25 and 298.67 GHz. Contours show the high–SNR region for cross-plot comparison. The CND has $\alpha \lesssim 1$, consistent with free–free emission, while the 20\kms{} and 50\kms{} molecular clouds have $\alpha \gtrsim 2$, characteristic of thermal dust emission. Right: Uncertainty on spectral index map with color bar clipped at $\Delta\alpha = 1$.}
        \label{fig:spectralindex}
    \end{figure*}

    The spectral index map (\autoref{fig:spectralindex}) shows that the continuum emission from the innermost parts of the CND has a relatively flat spectral index ($\alpha \lesssim 1$), consistent with free–free emission. In contrast, the 50~km\,s$^{-1}$ and 20~km\,s$^{-1}$ molecular clouds show $\alpha\gtrsim2$ across the bulk of these clouds.
    
    An alternate parameterization of the spectral variation of dust emission is through dust emissivity $\beta$, assuming the spectral profile is the blackbody emission profile multiplied by a power law with index $\beta$. With the assumption $T = 70$\,K (see~\autoref{appendix:h2estimationfromassumption}) and TIME's central frequency at 240 GHz, we fall into the Rayleigh-Jeans regime, where $\beta=\alpha-2$. In these terms, we find $\beta\gtrsim0$ in the 20\kms{} and 50\kms{} clouds (even $\beta\gtrsim2$ in core regions), clearly indicating (modified) blackbody emission from thermal dust.
    
    \subsection{Broadband Continuum Flux Map} \label{sec:broadband-flux-map}
    To verify our flux measurements and calibration, we compare an alternate broadband continuum flux map to data from the Bolocam Galactic Plane Survey (BGPS)~\citep{Ginsburg13}. We once again exclude frequency bands corresponding to the atmospheric monitoring channels, as described in~\autoref{sec:spectral_index}. However, for this comparison, we cannot average our per-frequency maps in a way that minimizes noise as described for computing our high SNR region, but instead must match the frequency bandpass of the BGPS observations (centered at \(270~\mathrm{GHz}\)). We thus collapsed each pixel's spectrum into a photometric flux in the BGPS effective observing band using spectral weights interpolated from the published Bolocam bandpass curve (e.g., as shown in~\citealt{BGPS1.0}).
    This alternate, BGPS bandpass-weighted continuum map should thus be as equivalent to BGPS as possible for flux comparison purposes. 

\begin{figure*}[t]
        \centering
        \includegraphics[width=0.486\linewidth]{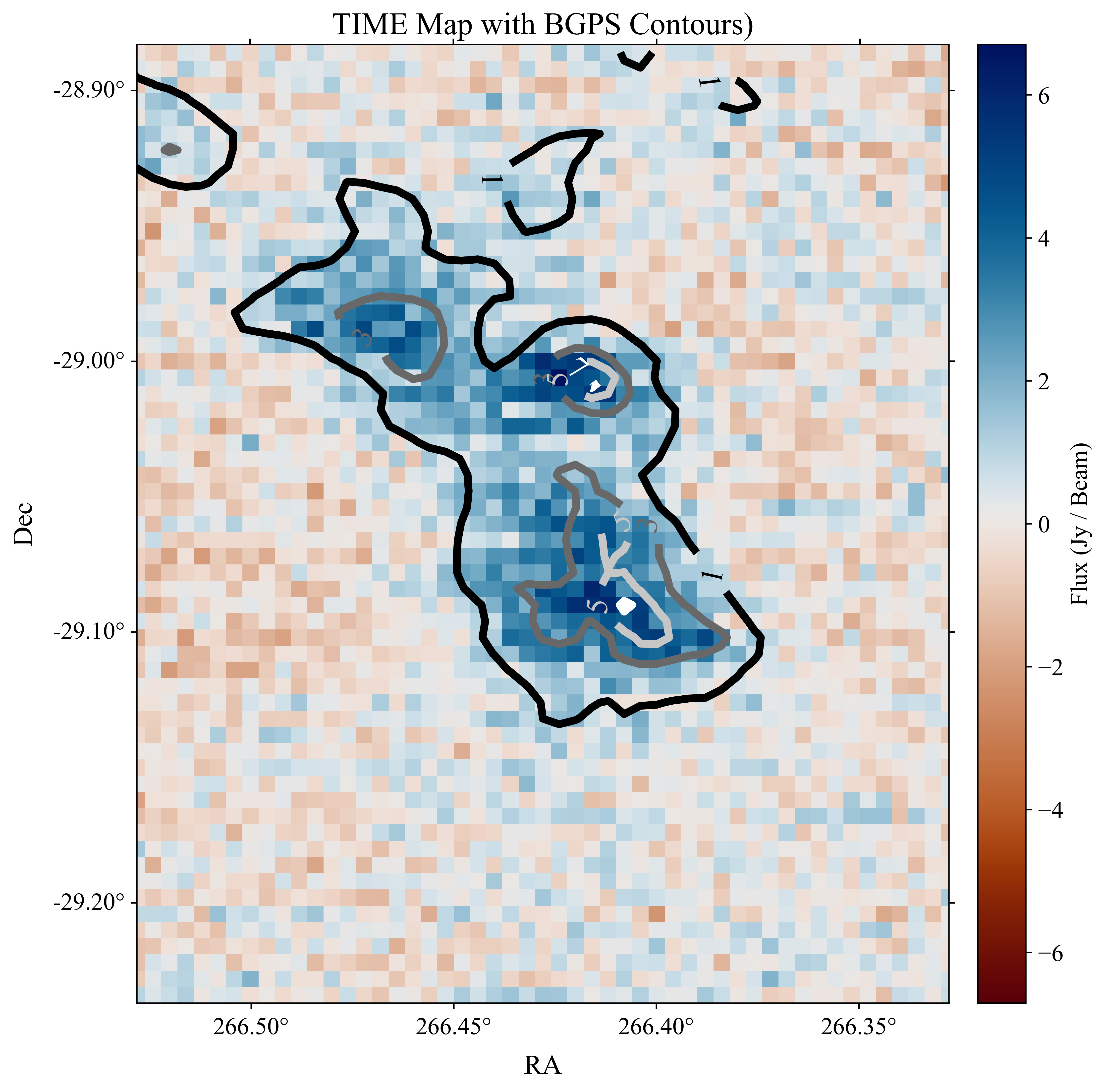}\includegraphics[width=0.486\linewidth]{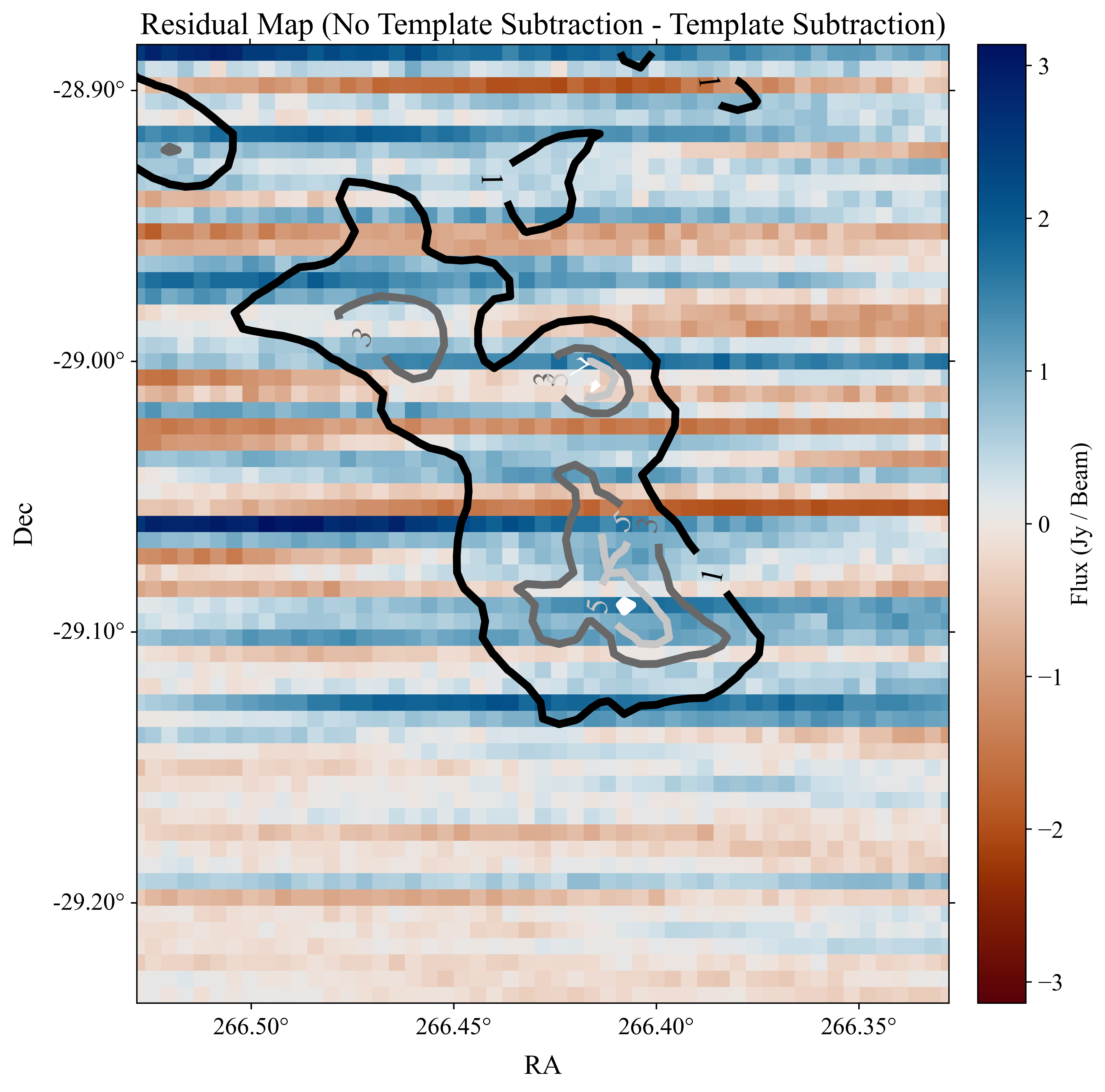}
        
        \includegraphics[width=0.486\linewidth]{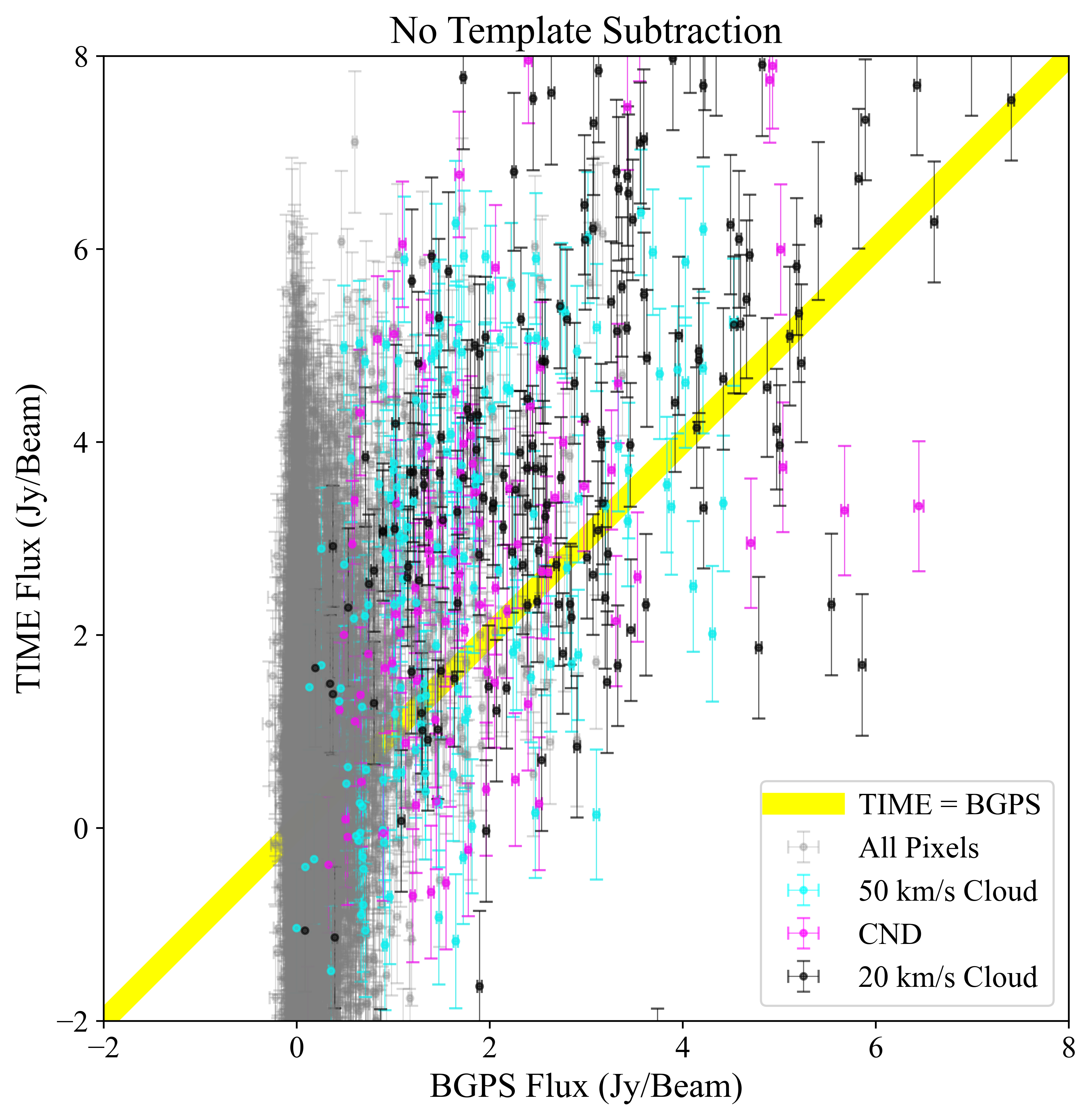}\includegraphics[width=0.486\linewidth]{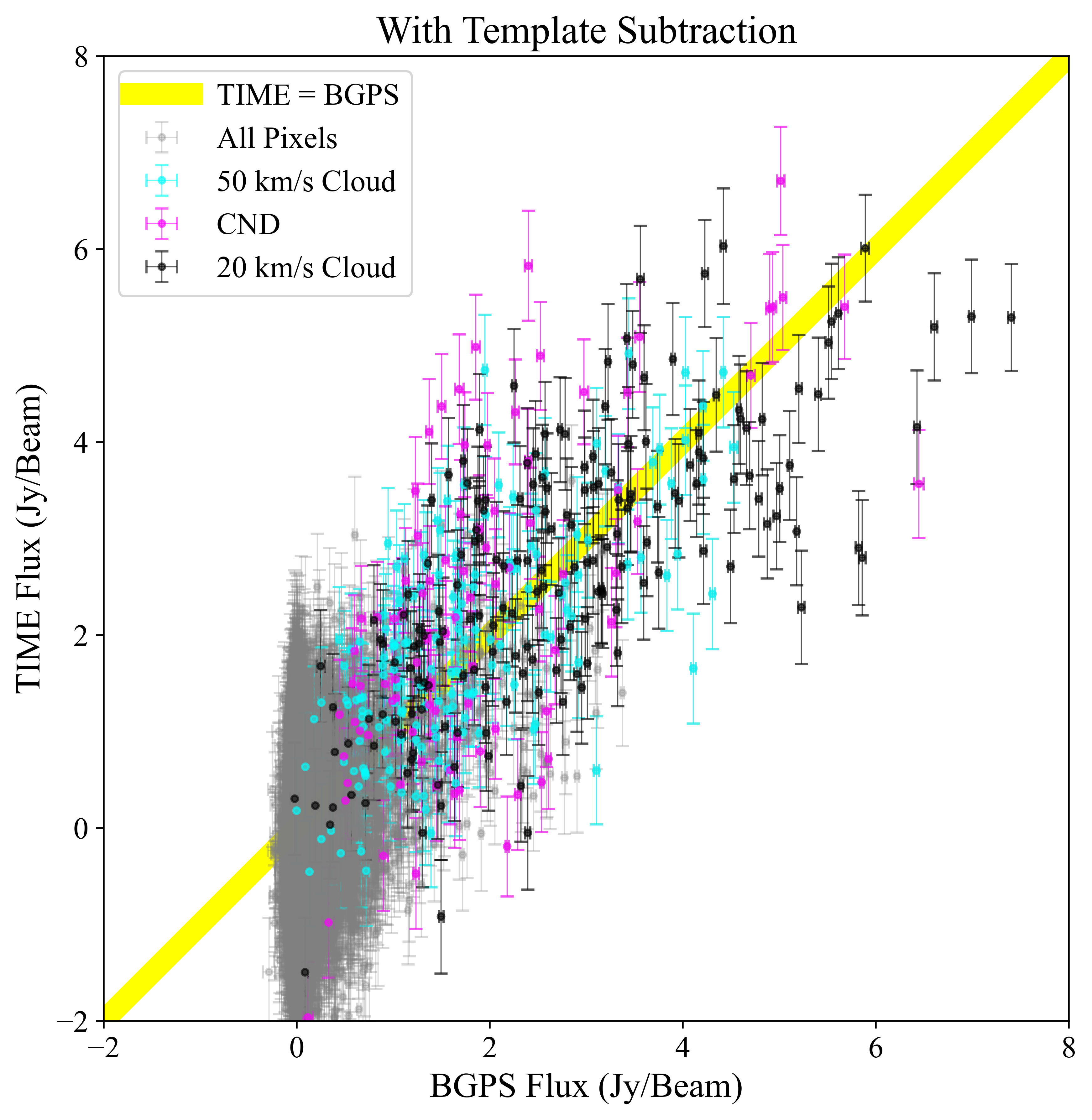}       
        \caption{Comparison of TIME broadband continuum flux map and BGPS's map of Sgr A. Top left: BGPS data reprojected to the TIME grid, and plotted as a contour over TIME's broadband continuum map computed from BGPS spectral weighting with contour levels 1, 3, 5, and 7. Top right: Residuals between broadband continuum flux maps produced without a common-mode template subtraction or PCA subtraction and with both subtractions. Bottom left: Pixel-wise comparison without a common-mode template or PCA subtraction from TIME's data. Scatter points are color-coded by spatial region, categorized according to the previously defined high-SNR sub-contours.
        A yellow 1–1 reference line provides an anchor for comparison. 
        Bottom right: The same plotting scheme as the bottom left plot, but now with both common-mode template and PCA subtraction applied to TIME data.}
        \label{fig:bgpscompare}
    \end{figure*}

    Before computing the pixel-pixel flux comparison, we first verify our spatial alignment with BGPS, as shown in the upper left panel of~\autoref{fig:bgpscompare}. Using Astropy \citep{robitaille2013astropy,price2018astropy,price2022astropy}, we transform BGPS's broadband continuum flux map into International Celestial Reference System (ICRS) coordinates to match TIME. We also corrected for the reported offset of BGPS (RA, Dec) of ($-2.16'', 2.52''$). The cleaned broadband map clearly aligns with the BGPS data, while the residuals from common-mode and PCA template subtraction clearly do not, as shown in the upper right panel of~\autoref{fig:bgpscompare}.

    Broadband continuum fluxes measured by TIME and BGPS mostly agree, as shown in the lower panels of~\autoref{fig:bgpscompare}. Template subtraction successfully removes correlated noise for improved agreement between BGPS and TIME flux measurements, both in the average and in the dispersion of TIME flux values around the BGPS measurements. Disagreement in pixels with higher surface brightness ($\gtrsim5$ Jy beam$^{-1}$) appears to arise from the time constants of the TIME detectors, which are as low as 10 ms in optimal conditions (well matched to the scan speed and sky pixelization), but can reach $\sim100$ ms under high loading and low loop gain with detectors near the top of the TES superconducting transition.

    While this dilutes TIME flux measurements in the brightest few pixels, it does not affect the excellent agreement in integrated broadband flux from each of the sources in the region. Both TIME and BGPS report a flux density of $120$ Jy from the 20\kms{} cloud; they respectively report $54$ Jy versus $48$ Jy from the CND, and $74$ Jy versus $67$ Jy from the 50\kms{} cloud. Across the entire high-SNR region, TIME reports $250$ Jy versus $240$ Jy calculated from the BGPS data. The integrated flux density obtained from TIME thus agrees with BGPS within 5\% overall from all sources detected in the TIME broadband map with high confidence, or within 2--12\% per source depending on the region. 
    
    We expect other observations will show even better pixel-to-pixel flux agreement, especially with more and better-tuned detectors in future observations, improving both the baseline level of correlated noise and our ability to identify and subtract noise templates. 
    
    \subsection{Integrated Spectra at Known Regions} \label{sec: integrated_spectra}
\begin{figure*}[hbt!]
        \centering
        \includegraphics[width=\linewidth]
        {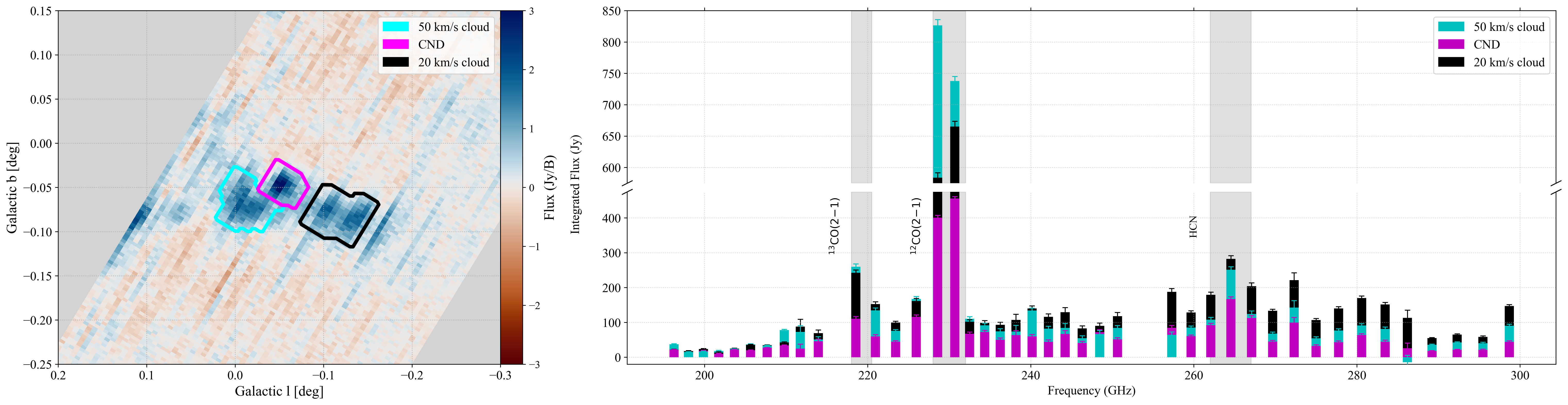}
        \caption{Left: Apertures (cyan, magenta, black) for the extraction of integrated spectra, overlaid on the continuum map used to compute the high-SNR pixels. Right: Spectra obtained in those apertures with error bars. Frequencies that correspond to the \xiiicotwoone{}, \xiicotwoone{}, and HCN transitions are shaded in gray.}
        \label{fig:broadband_spectra}
    \end{figure*}

     With understanding about our absolute flux calibration, we now extract integrated spectra for the $50$\kms{} cloud, the CND, and the $20$\kms{} cloud. We do this through aperture photometry: each position is defined by an aperture based on our high-SNR region, as shown in~\autoref{fig:broadband_spectra}. 
     Spectra are computed by integrating the flux within each aperture for all respective frequency channels without continuum subtraction, and converted to Jy through conversions dictated by our pixel area of $14.4'' \times 21.6''$ and the effective beam FWHM of $33''$. The uncertainty in each channel is the quadrature sum of the uncertainties in that channel for each pixel in the aperture.

    We see that the apparent widths of the CO lines are broader than the expected intrinsic values, most likely due to the finite spectral resolution of our observations and systematic deviations from the spectrometer design frequencies (see~\autoref{sec:instrumentbad} for additional discussion). However, the effective resolution achieved is still good enough to clearly distinguish the emission of the \xiicotwoone{} and \xiiicotwoone{} transitions. In addition, the C$^{18}$O(2--1) transition nominally overlaps with $^{13}$CO(2--1), but previous literature consistently suggests that the latter dominates the former by a factor of 5--10 in Sgr A and in similar environments \citep{Armstrong85,jimenez201713co,Ubagai19}.

\section{Discussion} \label{sec:discussion} 

    \subsection{Attributing Correlated Noise and Systematics to Sources in the TIME Instrument}
    \label{sec:instrumentbad}
    
\begin{figure*}
        \centering
        \includegraphics[width=\linewidth]{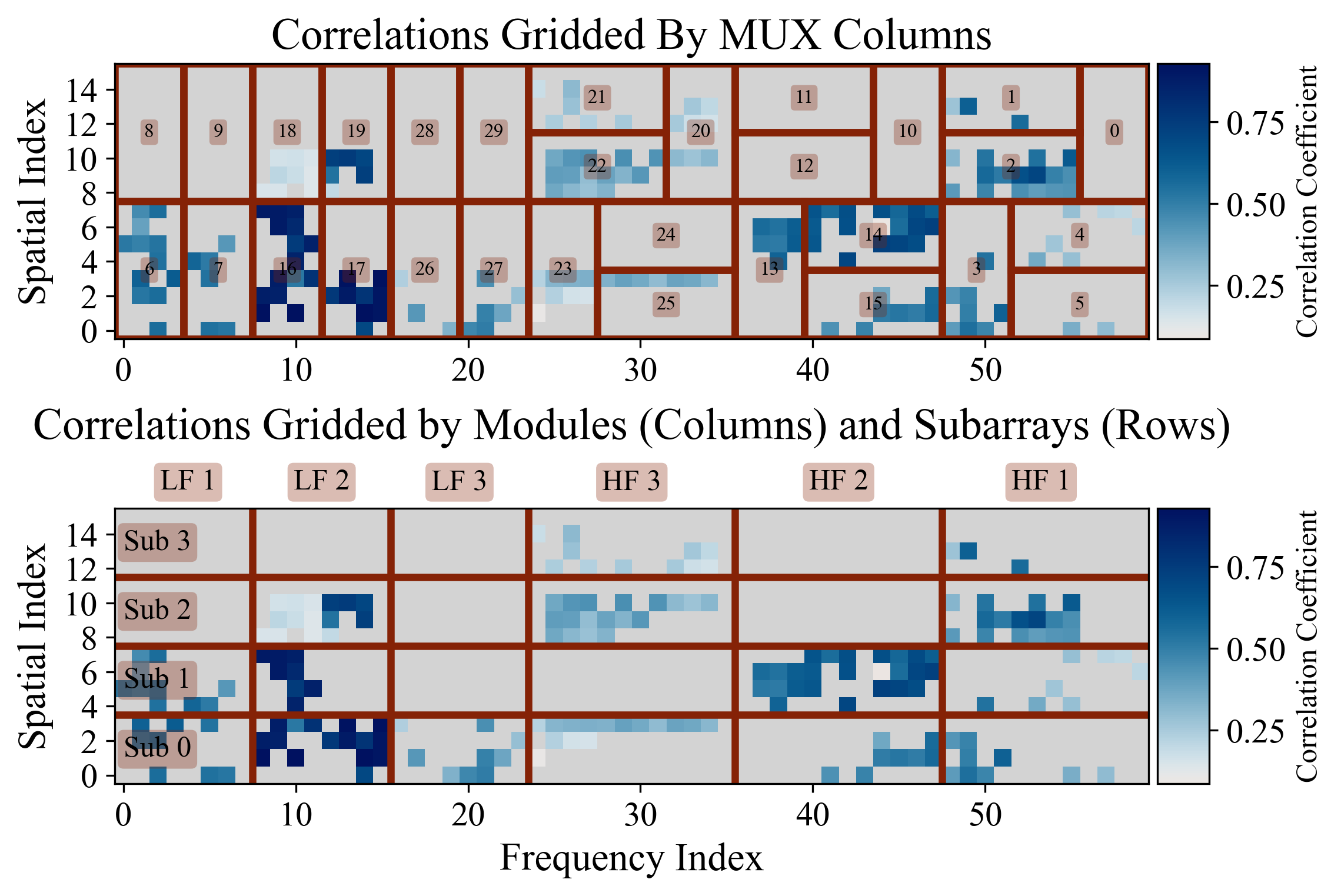}
        \caption{Correlation between individual detector TOD and their corresponding common–mode templates, plotted by mux-c (top) and by module/subarray (bottom). Structured patterns aligned with the detector layout suggest a strong association between the source of the correlated noise and the characteristics of the detector modules.}
        \label{fig:detector-correl-with-ts}
    \end{figure*}

    The common-mode template proved to be the most effective data-cleaning method in this analysis. This strong performance motivates a deeper investigation into the origin of the correlated noise. As an initial diagnostic, we plot the correlation coefficient between the middle 50\% of each detector TOD and the common-mode template to which that detector contributes. The spatial pattern of the correlation, shown in~\autoref{fig:detector-correl-with-ts}, shows clear structure when mapped against detector mux-c (top panel) and readout modules and subarrays (bottom panel), with particularly strong features in mux-c 16–19 and in the HF1 and HF2 modules. This detector-mapping aligned structure is an indication that the correlated noise originates from instrumental effects rather than astrophysical signals. 
    
    To better isolate the characteristic timescales of the correlated noise, we need to examine noise as a function of not just observing frequency and multiplexing readout, but also the TOD Fourier frequency. Large-scale atmospheric drifts would dominate noise at low Fourier frequency; instrumental effects tend to dominate noise at high Fourier frequency. We applied frequency-domain filtering using the low-pass ($<0.5$Hz), high-pass ($>5$Hz), and band-pass filters ($0.5-5$Hz).\footnote{Each filter uses a Gaussian roll-off with $\sigma=0.05$ at the band edges rather than a hard cut.} When applying each filter to all detector TOD and computing the pair-wise correlations in the same way as the observed all inter-detector correlation before common-mode template subtraction, shown in~\autoref{fig:correlation_of_detector}, we found that the 0.5--5\,Hz band-pass TOD correlation structure most clearly explains what we see in the all-Fourier correlation plots, contrary to the expectation of low-pass TOD correlations dominating in the case that atmospheric noise dominate. We quantify this by correlating the broadband matrix with those from each filtered band: the band-pass shows the strongest agreement ($r\simeq0.83$), followed by the high-pass ($r\simeq0.64$), and then the low-pass components ($r\simeq0.14$). 
    
    While common-mode subtraction is demonstrably effective, its very efficacy motivates the need to understand this correlated noise explicitly. We also aim to bring more good-quality detectors online in future observations. With a larger detector yield, the common-mode template can be estimated at finer granularity: per mux-c and per (module$\times$subarray) grid rather than coarse blocks. These templates would then capture instrument-originated correlated noise better while reducing leakage of sky modes. 
    
    In addition to the presence of correlated instrumental noise, our measurements of the CO lines indicate that the band centers of the detector frequency channels are not the exact predicted band centers. As will arise again in~\autoref{sec:co_line_analysis}, we expect the CO line profiles to be $\lesssim10^2$\kms{} wide, whereas the design channel bandwidth corresponds to a velocity width of $\sim10^3$\kms{}. Therefore, the CO line flux cannot possibly span multiple channels unless the central frequencies of the channels themselves are considerably offset. A subset of detector bandpasses has been characterized by an external Fourier-transform spectrometer (FTS). For example, lab tests using a different set of detector modules showed random variations of 0.5--1.0 GHz in the measured central or peak frequencies of frequency channels away from their design frequencies, and significant variation in the transmission profile. This may explain, for example, why the $^{13}$CO emission is less evident in the channel with the closest nominal central frequency (220.9 GHz) to the rest-frame frequency of 220.4 GHz, and more evident in the channel next to it (nominal frequency 218.6 GHz). Inspecting the limited lab data available for a different set of detectors suggested that the transmission profile for this lower-frequency channel did extend considerably towards the $^{13}$CO(2--1) emission frequency.
    
    These earlier tests only covered a subset of detectors in three feeds and therefore do not cover the focal plane in a way that usefully informs the present work. While precise band center and bandpass characterization are not needed for the above results, we will undertake full characterization of bands to inform future science with TIME (including LIM analyses) with the opto-mechanical design and focal plane configuration to be used for our science observations.

    \subsection{Analysis of CO Line-intensity Maps} \label{sec:co_line_analysis}

    We convincingly detect both \xiicotwoone{} and \xiiicotwoone{} line emission above continuum emission in the integrated spectra for the 50\kms{} and 20\kms{} clouds shown in~\autoref{fig:broadband_spectra}. Simultaneous observation of these isotopologues is interesting: \xiiico{} line emission is often optically thin, even when \xiico{} line emission is highly saturated and optically thick, allowing for a probe of physical conditions of the ISM. Indeed, one of the long-term objectives of astrophysics with LIM is to probe cosmic abundance ratios and saturation of such CO isotopologues, which in turn lends insights about star formation in high-redshift galaxies~\citep{breysse2017}.
    
    In the context of our Sgr A observation, we can demonstrate reliable relative flux calibration across our frequency channels by relating the observed $^{12}$CO$/^{13}$CO line ratio to the molecular hydrogen mass contained in the molecular clouds of this region, and comparing results to prior literature. We can even do this at a per-pixel level instead of in just the integrated spectra of~\autoref{sec: integrated_spectra}.

    We generate $^{12}$CO and $^{13}$CO flux maps by first using the spectral index fits for each pixel to subtract off the predicted continuum emission from our spectral data cube, propagating uncertainties in the continuum fitting into uncertainties in continuum-subtracted flux. As we discussed above in~\autoref{sec:instrumentbad}, it appears that the frequency channel nominally closest to the \xiiicotwoone{} emission frequency of $220.4$ GHz actually catches less of the line flux than the channel just below it in frequency (nominal central frequency 218.55 GHz). Therefore, we use the flux measured in that channel as our best estimate of the \xiiicotwoone{} line flux. However, the two frequency channels closest to the \xiicotwoone{} emission frequency of $230.5$ GHz catch excess flux above continuum in roughly equal measure, so we take the inverse-variance weighted average of those two channels (nominal central frequencies 228.58 GHz and 230.68 GHz) to estimate the \xiicotwoone{} flux.
    
    We average rather than sum across these channels because our expectations for the intrinsic velocity width of the spectral profile are 1--2 orders of magnitude narrower than the expected channel bandwidth. Therefore, the two channels capturing \xiicotwoone{} emission are best treated as independent measurements of the same line flux, rather than sampling different parts of the line profile. In fact, in the case of \xiiicotwoone{}, it is likely that the line flux is sampling different parts of the transmission profiles of overlapping channels. We survey values of intrinsic velocity width in~\autoref{appendix:h2estimationfromassumption}, in the context of CO excitation calculations.

    We also compare our $^{13}$CO flux maps to data from the SEDiGISM survey~\citep{schuller2020sedigism}. The spatial emission features observed by TIME closely track those from SEDiGISM, and the velocity-integrated fluxes from the individual sources are in remarkable agreement. From the 50\kms{} cloud, TIME measures $5.9\times10^5$ Jy\kms{} compared to SEDiGISM's $5.3\times10^5$ Jy\kms{}; from the 20\kms{} cloud, TIME measures $5.4\times10^5$ Jy\kms{} compared to SEDiGISM's $5.1\times10^5$ Jy\kms{}; and from the CND, TIME measures $2.3\times10^5$ Jy\kms{} compared to SEDiGISM's $2.1\times10^5$ Jy\kms{}. This excellent agreement within $\approx10$\% gives us confidence that the measurement of excess flux above continuum truly reflects the CO line flux. A visual comparison between the SEDiGISM and TIME \(^{13}\mathrm{CO}\) maps suggests that SEDiGISM resolves finer substructure within the 50\kms{} cloud, again likely due to TIME detectors operating with high optical load as with the broadband flux comparison in~\autoref{sec:broadband-flux-map}. 

    For each pixel, the $^{12}$CO$/^{13}$CO line ratio is the ratio of the $^{12}$CO and $^{13}$CO continuum-subtracted, averaged flux maps described above. We excluded pixels where the uncertainty in the line ratio exceeded ten times the computed ratio.

    We derive a map of total CO column density based on the line ratio map, according to a calculation that assumes local thermodynamic equilibrium (LTE) and assumes specific values for physical factors based on prior literature. We detail this calculation and the underlying assumptions in~\autoref{appendix:h2estimationfromassumption}.

    Assuming a [CO]/[H$_2$] ratio of $10^{-4}$ \citep{vandishoeck1987}, we are able to then estimate the total mass of molecular hydrogen contained in the 50\kms{} cloud, the CND, the 20\kms{} cloud, and the entire region of high SNR pixels. We tabulate these masses in~\autoref{tab:h2mass_subtotals}.

    We may also estimate the H$_2$ mass from the $^{12}$CO flux by assuming a conversion $X_\text{CO}$ between the integrated $^{12}$CO brightness temperature map and H$_2$ column density. Based on \cite{kohnosofue24}, we use a Galactic value of $X_{\text{CO}}=2.2 \times 10^{20}$ cm$^{-2}$ $(\text{K km s}^{-1})^{-1}$, the resulting mass estimates are slightly higher but within an order of magnitude, finding a total mass of $5.7\times10^5\,M_\odot$ across the high-SNR regions. We discuss this further in~\autoref{appendix:h2estimationfromassumption}.

    The total mass obtained is broadly in line with values reported in the literature. For comparison, \citet{oka1998} estimated the mass of the Sgr A complex to be $4.5 \times 10^{4}$ to $1.1 \times 10^{5}$ $M_{\odot}$ under the assumption that the clouds are in pressure equilibrium, and $5.4 \times 10^6$ to $1.1 \times 10^8$ $M_{\odot}$ if the clouds are instead gravitationally bound. More recently, \cite{3DCMZ-I,3DCMZ-II} analyzed \emph{Herschel} data to constrain physical dust temperature and H$_2$ column density across the entire CMZ, subsequently undertaking a dendrogram-based analysis to place molecular clouds in a hierarchical catalog. This analysis found a total H$_2$ mass of $5.5\times10^5\,M_\odot$ for the combination of the 20\kms{} and 50\kms{} clouds, bracketed nicely by our estimates of the mass summed across the high-SNR regions: $5.4\times10^5\,M_\odot$ and $5.7\times10^5\,M_\odot$ from the line ratio and $X_{\text{CO}}$ computations respectively. 

    \begin{deluxetable}{lc}
        \tablecaption{H$_2$ mass of regions within the TIME observation of Sgr A, estimated from the $^{13}$CO/$^{12}$CO ratio.
        \label{tab:h2mass_subtotals}}
        \tablehead{
        \colhead{Region\tablenotemark{a}} & \colhead{H$_2$ mass ($M_\odot$)}}
        \startdata
        50\kms{} cloud & $1.9\times10^5$\\
        CND & $6.8 \times10^4$\\
        20\kms{} cloud & $2.9 \times10^5$\\\hline
        {\bf High-SNR regions} & $5.4\times10^5$\\\hline
        Total across map & $3.4\times10^6$ 
        \enddata
        \tablenotetext{a}{{We show boundaries for these regions in~\autoref{fig:spectralindex} and ~\autoref{fig:broadband_spectra}.}}
    \end{deluxetable}

\section{Conclusions}

In this work, we present the first commissioning observations of the TIME instrument using the ARO 12-meter telescope at Kitt Peak. Our analysis demonstrates that TIME can successfully acquire and process broadband millimeter-wave spectral maps of complex astrophysical regions, even under the high-noise conditions of early engineering runs and low elevations. 

\paragraph{Calibration and map fidelity.} A planet-based calibration using \textit{SPACETIME} produces frequency-dependent gains and feed offsets that, when applied to Sgr~A, recover: (i) a continuum spectral index map that is flat in the CND (free–free; $\alpha\lesssim1$) and dust-like in the $20$ and $50$\kms{} clouds ($\alpha\gtrsim2$), and (ii) strong detections of $^{12}$CO(2--1) and $^{13}$CO(2--1). After BGPS spectral bandpass correction, broadband continuum fluxes agree with BGPS at the $\sim$5\% level in TIME's overall high SNR regions. 

\paragraph{Correlated noise} The dominant systematic is a correlated component aligned with detector mapping, peaking at Fourier frequencies between 0.5\,Hz to 5\,Hz. A two-tier, correlation-weighted common-mode subtraction by mux-c and frequency blocks is used to clean this detector-correlated noise. Across the whole map, this correlated noise removal reduces flux by a factor of $\sim2$, and it was an acceptable trade in this high-noise low-elevation dataset. We followed this with a PCA-based cleaning of any map-space systematics. Several components that weakly correlated with the signal were removed without obvious over-subtraction of the source structure. 

\paragraph{Spectral and astrophysical inferences.} 
Strong $^{12}$CO(2–1) and $^{13}$CO(2–1) emission is detected toward the bright molecular clouds surrounding Sgr~A. We are able to derive molecular hydrogen mass estimates using maps of the $^{12}$CO and $^{13}$CO line fluxes, with results falling well within the expected range reported from the literature.

The work also defines priorities for our future science developments: (a) increase detector yield to enable per–mux-c and per–(module$\times$subarray) templates with less sky leakage, and (b) refine noise modeling in the 0.5–5\,Hz regime to reduce reliance on aggressive cleaning that compromises flux recovery.

\begin{acknowledgements}
This material is based upon work supported by the National Science Foundation under Grant Nos.~1602677, 1653228, 1910598, 2308039, 2308040, 2308041, and 2308042.

The TIME collaboration acknowledges the critical contributions of the Arizona Radio Observatory engineers and operators, including Kevin Bays, Tom Folkers, Natalie Gandilo, Blythe Guvenen, Sean Keel, and Martin McColl. We also acknowledge Sukhman Singh and Ibrahim Shehzad at Cornell and Lisa Nasu-Yu at the University of Toronto for early contributions to data analysis, observations, and simulations that ultimately informed this work.

SMM acknowledges support as a Nexus Scholar from the Cornell University College of Arts \& Sciences.

ATC acknowledges support as a Fred Young Faculty Fellow at Cornell University.

DTC was supported by a CITA/Dunlap Institute postdoctoral fellowship for much of this work. The Dunlap Institute is funded through an endowment established by the David Dunlap family and the University of Toronto. Research in Canada is supported by NSERC and CIFAR.

This publication is based on data acquired with the Atacama Pathfinder Experiment (APEX) under programmes 092.F-9315 and 193.C-0584. APEX is a collaboration among the Max-Planck-Institut fur Radioastronomie, the European Southern Observatory, and the Onsala Space Observatory. The processed data products are available from the SEDiGISM survey database located at https://sedigism.mpifr-bonn.mpg.de/index.html, which was constructed by James Urquhart and hosted by the Max Planck Institute for Radio Astronomy.

\end{acknowledgements}

\begin{contribution}

SFY led the bulk of the data analysis, including TOD processing and continuum analysis, and supervised SMM's contributions. SMM implemented PCA of the maps and led the analysis and scientific interpretation of line emission present in the reduced maps. BJV contributed software for calibrator observation processing. Principal investigator ATC led the acquisition of funding to support the project, supervised on-site instrument operations and remote observing, and oversaw progress on the project and the writing of this article.

Authors 5--8 contributed towards this work in a previous form, including data acquisition and earlier iterations of data analysis. 

All other authors are alphabetically listed members of the TIME collaboration, and contributed scientific and technical expertise directly relevant to the analysis and interpretation presented throughout this work.

\end{contribution}
\appendix

\section{Common-Mode Template Supplement} \label{appendix:cm-template-supplement}

A common-mode template is an estimate of the TOD component shared across multiple detectors in an array, constructed to capture patterns in the TOD that arise from correlated noise rather than true astrophysical emission. In our case, we identify those by taking the median across detector blocks at each time sample.

\begin{figure}
        \centering
        
        \begin{subfigure}{\linewidth}
            \centering
            \includegraphics[width=\linewidth]{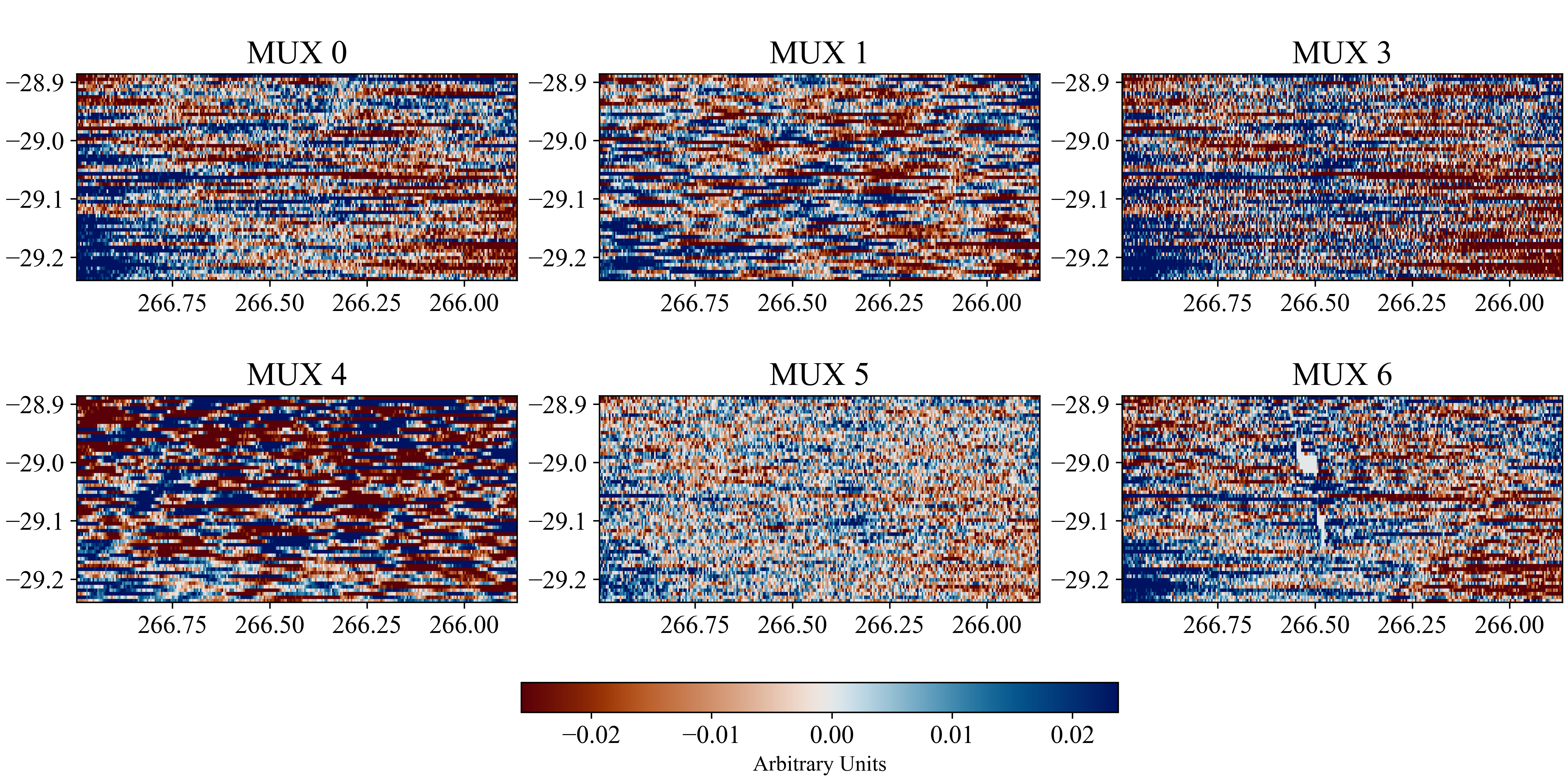}
            \caption{The final mux-c templates fold directly into the maps, excluding templates \#2 and \#7 because no detectors contribute to them. All three detectors contributing to template \#6 show high correlation and are flagged, which creates the empty regions where signal would otherwise have leaked.}
            \label{fig:muxround3}
        \end{subfigure}
        
        \vspace{1em} 
        
        \begin{subfigure}{\linewidth}
            \centering
            \includegraphics[width=\linewidth]{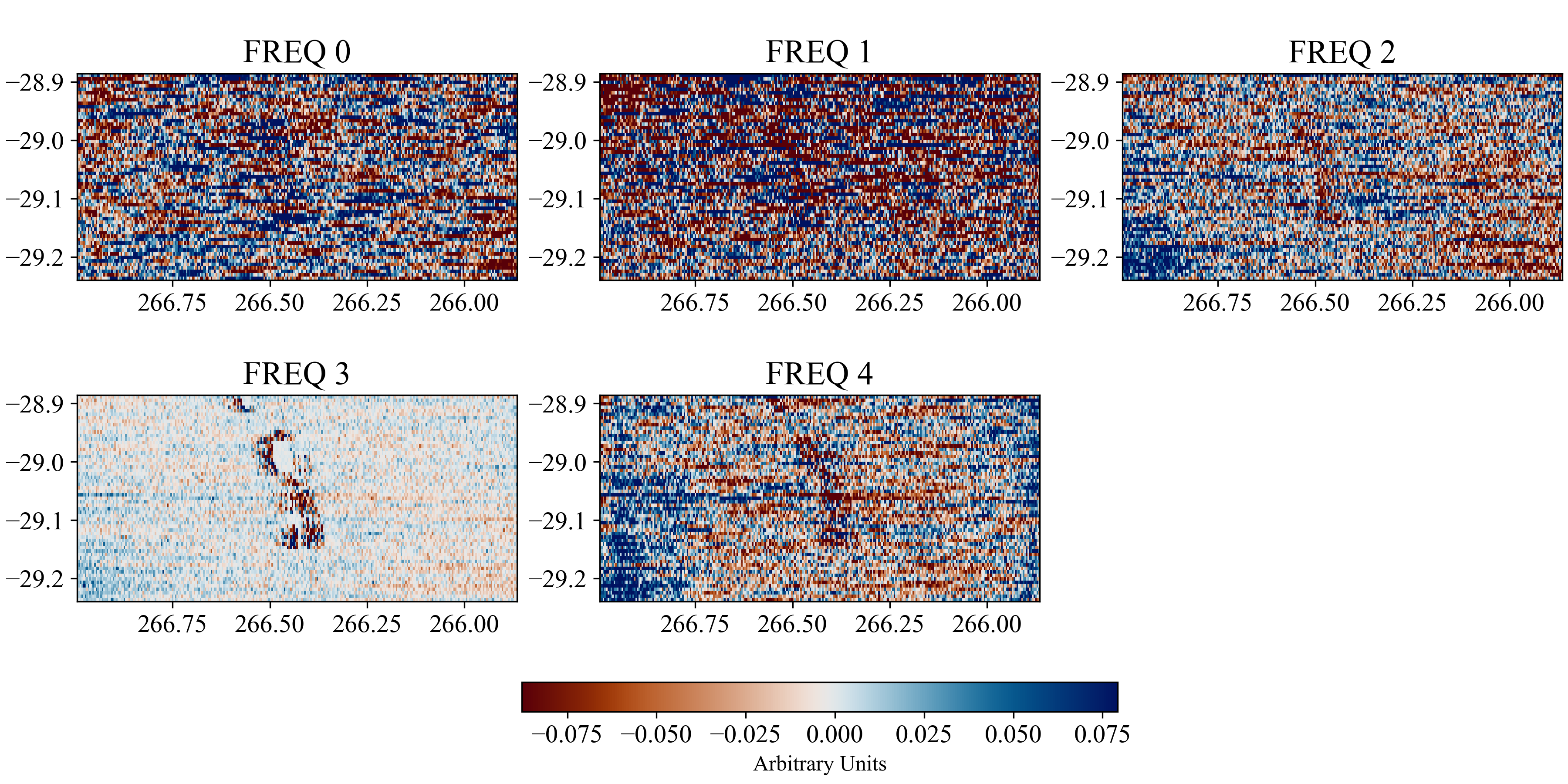}
            \caption{The final frequency templates fold directly into the maps. Although they still contain some signal structure, their subtraction is far less aggressive than the mux-c common-mode subtraction, and PCA was able to identify this structure in later components. Therefore, we avoid further iterative cleaning to prevent over-subtraction.}
            \label{fig:freqround3}
        \end{subfigure}
        
        \caption{Comparison of final common-mode template subtractions: (a) mux-c templates and (b) frequency templates.}
        \label{fig:commonmode}
    \end{figure}

Here, in~\autoref{fig:commonmode}, we fold the final set of common-mode templates into maps, which show the spatial structures these correlated signals would imprint onto our flux maps if not subtracted away. We show both mux-c and frequency block templates, and refer the reader to~\autoref{sec:common-mode-noise} for further discussion of the block definition and template calculation.

PCA played a critical role in diagnosing artifacts in these common-mode templates. In an earlier iteration of this analysis, several principal components exhibited shadow structures resembling the Sgr A cloud complex, but not at the correct spatial coordinates at first. Upon careful visual inspection, we confirmed these patterns were also present at low levels in the individual detector maps. We traced these shadows back to the common-mode subtraction step and evolved our template construction procedure into an iterative process accordingly. 

\section{H$_2$ estimation details and differences from various assumptions} \label{appendix:h2estimationfromassumption}

\begin{figure}
    \centering
    \includegraphics[width=\linewidth]{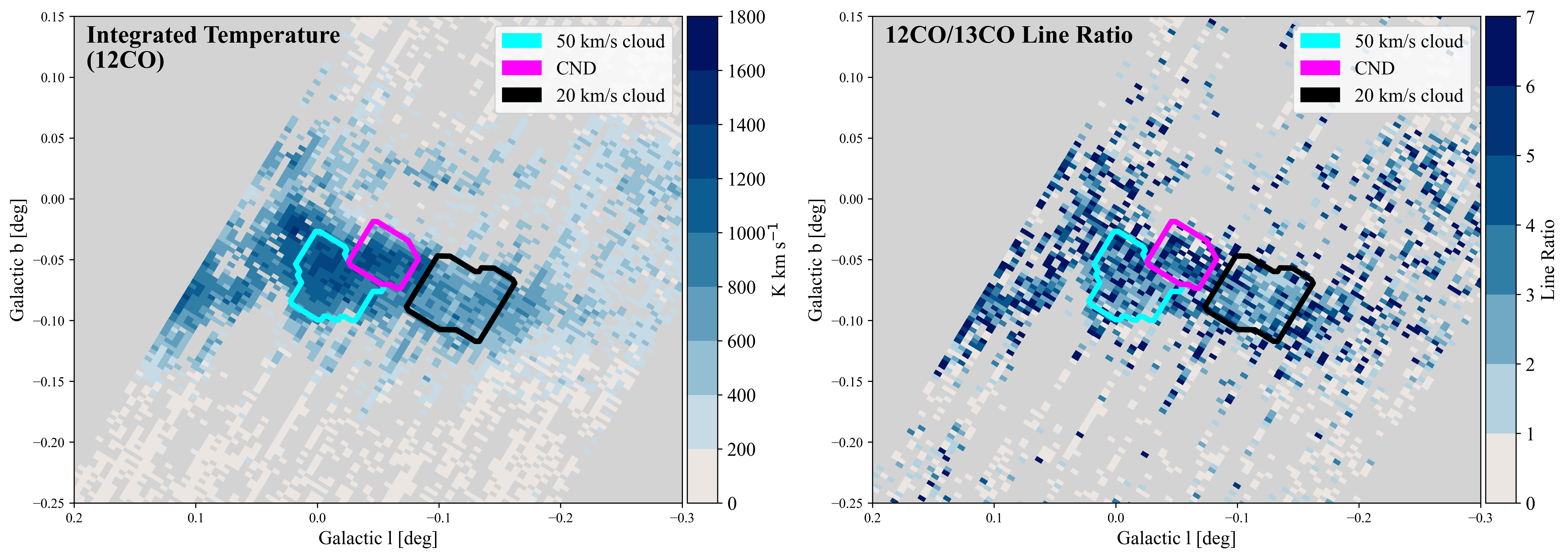}
    \caption{Maps of the $^{12}$CO velocity-integrated brightness temperature \emph{(left)} and $^{12}$CO/$^{13}$CO line ratio \emph{(right)} across the Galactic center region derived from TIME data.}
    \label{fig:appendix_maps}
\end{figure}

The TIME data allows derivation of H$_2$ mass from either the $^{12}$CO/$^{13}$CO line ratio (via CO and H$_2$ column densities) or the $^{12}$CO temperature alone (via H$_2$ column density), both of which we map out across the Sgr A region as shown in~\autoref{fig:appendix_maps}.

The derivation of total CO column density from the $^{12}$CO/$^{13}$CO line ratio is as follows. First, we relate the column density of $^{12}$CO in the lower state, $N^{l}_{12}$, to the observed line ratio $I_{13}/I_{12}$ and the isotopologue abundance ratio $R = [^{13}\mathrm{CO}]/[^{12}\mathrm{CO}]$:
    \begin{equation}
            \frac{I_{13}}{I_{12}} = \left(\frac{\nu_{13}}{\nu_{12}}\right)^2\left(\frac{A_{13}}{A_{12}}\right)\left(\frac{\sigma_{12}}{\sigma_{13}}\right)\left(e^{-h(\nu_{13}-\nu_{12})/k_{B}T_{ex}}\right)\times\frac{1-\text{exp}(-R\sigma_{13}N^l_{r12})}{1-\text{exp}(-\sigma_{12}N^l_{r12})},
    \end{equation}
    where $I_{13}$ and $I_{12}$ are the observed brightness of $^{13}$CO and $^{12}$CO, respectively, and $T_\text{ex}$ is the excitation temperature describing the relative occupation of the different rotational energy levels of CO. The Einstein A coefficients, $A_{12}$ and $A_{13}$, for the $J=2\to1$ transition of $^{12}$CO and $^{13}$CO are $6.91 \times 10^{-7}$ s$^{-1}$ and $6.08 \times 10^{-7}$ s$^{-1}$ respectively \citep{endres2016,muller2001,muller2005}. $\sigma_{13}$ and $\sigma_{12}$ are cross-sections of $^{13}$CO and $^{12}$CO corrected for stimulated emission, respectively, and are given by:
    \begin{equation}
        \sigma = \frac{3c^2A_{ul}}{8\pi\nu^2\Delta\nu_{\text{FWHM}}}\left(1-e^{-h\nu/k_bT_{\text{ex}}}\right).\label{eq:breysse2017}
    \end{equation}
  
    We assume LTE and therefore consider $T_\text{ex}$ to be equal to the gas kinetic temperature.
    
    The value of $R$ (the abundance ratio of $^{13}$CO to $^{12}$CO) in the Galactic center is approximately 1/24 \citep{langer1990}. Additionally, previous studies have found that the gas temperature in the Central Molecular Zone of the Milky Way is approximately 50--100 K \citep{Henshaw23}, and that the Galactic center molecular clouds have a velocity FWHM of 20--50\kms{} \citep{bally1987}. We thus assumed an excitation temperature of 70 K and a $\Delta\nu_\text{FWHM}=35$\kms{}, although we also explore the effect of varying the assumed excitation temperature in~\autoref{appendix:h2estimationfromassumption}. With these assumptions, we were able to use~\autoref{eq:breysse2017} to calculate the column density of $^{12}$CO in the lower state based on our observed line ratios. Pixels where Scipy's fsolve failed to find an adequate solution (which were largely off-source) were excluded from subsequent mass calculations.

    To determine the total CO column density, we need to calculate the fraction of CO molecules in the upper $J = 1$ state. By combining equations 31, 33--35, 51, and 53 from \citep{mangum2015}, the fraction of CO molecules in energy level $J_u$ is given by

    \begin{equation}
        \frac{N_u}{N_{tot}}=\frac{hB_0(2J_u+1)}{k_BT}e^{-hB_0[J_u(J_u+1)+\frac{1}{3}]/k_BT}
    \end{equation}
    Using values of $B_0\approx\nu/2$ from~\cite{nistdiatomic}, we find the fraction of $^{12}$CO that is in the $J = 1$ state under our assumed Galactic center conditions is 0.108, and the fraction of $^{13}$CO is 0.104. As expected, most of the CO molecules should be in the ground state, and the correction from including molecules occupying the upper rotational energy state is relatively small.

Since the H$_2$ mass of Sgr A then derives from the total CO column density, the mass determination depends on the assumed excitation temperature for the Galactic Center molecular clouds. The estimated masses across the temperature range of 50--100 K quoted by \cite{Henshaw23} are displayed in~\autoref{tab:h2mass_tempassumptions}; we used 70 K in the main line ratio analysis as it is in the center of this range. Overall, assuming a hotter excitation temperature increases the estimated H$_2$ mass. In addition, the estimated mass at 100 K differs from the estimated mass at 50 K by a factor of $3.5$ for each region.

For completeness, we also use~\autoref{tab:h2mass_tempassumptions} to show the masses calculated from the $^{12}$CO fluxes alone using an assumed $X_\text{CO}$ conversion. Our specific assumed $X_\text{CO}$ value derives from the average across all Galactic giant molecular clouds surveyed in the work of \cite{kohnosofue24}. However, the same work finds some evidence for decreasing $X_\text{CO}$ with lower galactocentric distance; extrapolating their results would suggest $X_{\text{CO}}=1.6 \times 10^{20}$ cm$^{-2}$ $(\text{K km s}^{-1})^{-1}$, for H$_2$ masses lower by a factor of 1.4. The work, although the most recent source for an estimate of $X_\text{CO}$ within the inner 8 kpc of the Galaxy, is also discrepant with some prior studies of $X_\text{CO}$ near the Galactic center. \cite{kohnosofue24} note that their result is higher by a factor of 3 than that of \cite{Arimoto96}, for example. It is possible, therefore, that our assumed $X_\text{CO}$ value, and therefore the corresponding mass estimates from $^{12}$CO flux alone, should be adjusted down by a factor of 2--4. But such an adjustment seems necessary only in the vicinity of the CND, and even this may be a natural consequence of our extremely approximate assumptions. This includes not least the assumption of LTE, which almost certainly does not hold in the immediate vicinity of Sgr A$^*$.

\begin{deluxetable}{lccccccc}
    \tablecaption{Estimated H$_2$ mass ($M_\odot$) of Sgr A regions computed from $^{12}$CO/$^{13}$CO line ratios for various assumed excitation temperatures, or alternatively using only the $^{12}$CO flux. We indicate the fiducial value of 70 K in bold type.
    \label{tab:h2mass_tempassumptions}}
    \tablehead{
   \colhead{Region\tablenotemark{a}} & \colhead{50 K} & \colhead{60 K} & \colhead{\bf 70 K} & \colhead{80 K} & \colhead{90 K} & \colhead{100 K} & \colhead{$^{12}$CO only\tablenotemark{b}}}
   \startdata
   50\kms{} cloud & $1.0 \times10^5$ & $1.4\times10^5$ & $1.9  \times10^5$ & $2.4\times10^5$ & $3.0 \times10^5$ & $3.6 \times10^5$& $2.5 \times 10^5$\\
   20\kms{} cloud & $1.6\times10^5$ & $2.2\times10^5$ & $2.9 \times10^5$ & $3.7 \times10^5$ & $4.6\times10^5$ & $5.6\times10^5$ & $1.9 \times 10^5$ \\
   CND & $3.7\times10^4$ & $5.1 \times10^4$ & $6.8 \times10^4$ & $8.6 \times10^4$ & $1.1 \times10^5$ & $1.3 \times10^5$ & $1.2 \times 10^5$ \\
   \hline
   {\bf High-SNR regions} & $2.9  \times10^5$ & $4.1 \times10^5$ & $5.4 \times 10^5$ & $6.9 \times10^5$ & $8.6 \times10^5$ & $1.1 \times10^6$ & $5.7 \times10^5$\\\hline
   Total across map & $1.8 \times10^6$ & $2.5\times10^6$ & $3.4 \times10^6$ & $4.3 \times10^6$ & $5.4 \times10^6$ & $6.5 \times10^6$ & $1.5\times 10^6$\\
   \enddata
   \tablenotetext{a}{{These are the same regions as in~\autoref{tab:h2mass_subtotals}.}}
   \tablenotetext{b}{{These estimates derive from the velocity-integrated $^{12}$CO brightness temperature assuming $X_{\text{CO}}=2.2 \times 10^{20}$ cm$^{-2}$ $(\text{K km s}^{-1})^{-1}$.}}
\end{deluxetable}

\bibliography{biblio}
\bibliographystyle{aasjournalv7}
\end{document}